\documentclass{aa}  
\usepackage[colorlinks=true, linkcolor=blue, citecolor=blue, urlcolor=blue]{hyperref}
\usepackage{graphicx}
\usepackage{txfonts}
\usepackage{subcaption}
\usepackage{natbib}
\bibpunct{(}{)}{;}{a}{}{,} 

\begin{document}

    \title{Relative alignment between gas structures and magnetic field in Orion A at different scales using different molecular gas tracers}

   \authorrunning{Jiao, Wang et al. }

   \author{Wenyu Jiao
          \inst{1,2,3}
          \and Ke Wang \inst{2}
          \and Fengwei Xu \inst{1,2}
          \and Chao Wang \inst{4}
          \and Henrik Beuther \inst{3}
          }

   \institute{Department of Astronomy, School of Physics, Peking University, Beijing, 100871, People’s Republic of China 
   \and 
   Kavli Institute for Astronomy and Astrophysics, Peking University, Beijing 100871, People’s Republic of China \\ \email{kwang.astro@pku.edu.cn}  \and Max Planck Institute for Astronomy, Königstuhl 17, D-69117 Heidelberg, Germany 
         \and
             National Astronomical Observatories, Chinese Academy of Sciences, 20A Datun Road, Chaoyang District, Beijing 100101, China \\
             }

   \date{Received 9 January 2024; accepted xx xx, 2024}

 
  \abstract
   {Magnetic fields can play a crucial role in high-mass star formation. Nonetheless, the significance of magnetic fields at various scales and their relationship with gas structures have been largely overlooked.}
   {Our goal is to examine the relationship between the magnetic field and molecular gas structures within the Orion A giant molecular cloud at different scales and density regimes.}
   {We assessed the gas intensity structures and column densities in Orion A using $^{12}$CO, $^{13}$CO, and C$^{18}$O from Nobeyama observations. By comparing Nobeyama observations with {\it{Planck}} polarization observations on large scales ($\sim0.6$ pc) and JCMT polarization observations on small scales ($\sim0.04$ pc), we investigate how the role of magnetic fields changes with scale and density.}
   {We find a similar trend from parallel to perpendicular alignment with increasing column density in Orion A at both large and small spatial scales. In addition, when changing from low-density to high-density tracers, the relative orientation preference changes from random to perpendicular. The self-similar results at different scales indicate that magnetic fields are dynamically important in both cloud formation and filament formation. However, magnetic field properties at small scales are relative complicated, and the interplay between magnetic field and star-forming activity needs to be discussed case by case. }
   {}

   \keywords{Molecular cloud, high-mass star formation, magnetic fields,  Polarimetry
               }
   \titlerunning{Alignment between gas structures and magnetic field in Orion A}
   \maketitle





\section{Introduction} \label{sec:intro}
The formation and evolution of molecular clouds are intricate processes driven by the interplay of several crucial factors, which include gravity, turbulence, magnetic fields, thermal instability, feedback, and cosmic-ray ionization. These factors play different roles in different activities, such as filamentary structure formation and star formation processes \citep[e.g.,][]{2007ARA&A..45..339B,2007ARA&A..45..565M,WangK2015book}. The magnetic field is considered as one of the key components in slowing down the star formation rate, but it is difficult to measure the magnetic field because of 
observational limitations \citep{2012A&ARv..20...55H}.

Until now, two primary methods have been used to measure the magnetic field in molecular clouds. Observations of the Zeeman effect are the only direct way to measure the strength of the magnetic field in interstellar clouds. Some observations have  succeeded in 
calculating the value of the magnetic field along the line of sight \citep[e.g.,][]{1996ApJ...456..217C,2008_CN_Zeeman_A&A...487..247F}. However, because the Doppler broadening effect is always larger than the Zeeman splitting effect, this method is limited by observational techniques and cannot be widely used \citep{2012_Crutcher_ARA&A..50...29C}.

An alternative way to infer the morphology of magnetic fields is through observations of linear polarization induced by interstellar dust. One possible and longstanding method is to measure the polarization from the background starlight, which assumes aspherical 
dust grains align with their long axis perpendicular to the local magnetic field and produce polarization parallel to the field projected on the plane of sky (see \citealt{Anderson_2015_review} for a review). However, it is difficult to fully reproduce the projected magnetic field morphology inside molecular clouds. In recent years, a more general method has been to measure the orientation of magnetic fields within the molecular clouds through the measurement of linearly polarized radiation emitted from dust, which can map the morphology of magnetic fields on the plane of sky on small scales around the star-forming regions \citep[e.g.,][]{Matthews_2009, Zhang_2014} or on large scales across the whole sky \citep{planck_2020a}, helping us to gather more information about magnetic fields.

The strength of the magnetic field projected on the plane of sky can be estimated using the Davis-Chandrasekhar-Fermi method \citep{Davis,CF}, but the uncertainty and applicability of the DCF method are hotly debated \citep[e.g.,][]{Liu_2021, Skalidis_2021a,Skalidis_2021b, Chen_2022}. Some studies have sought to derive additional information about the role of magnetic field by comparing the relative orientation of magnetic field inferred from dust polarization with the orientation of molecular cloud structures, and have revealed a strong correlation between magnetic field morphology and the geometry of the molecular cloud structures \citep[e.g.,][]{Tassis_2009,Koch_2013, planck_HRO_diffuse, Planck_2016c}. Multiscale magnetic fields have also been explored by various authors. These studies show that the role of the magnetic field may vary at different scales  \citep[e.g.,][]{Li_2015, Chen_2020,Liu_2023}.

In recent years, the histograms of relative orientation (HRO) technique was introduced to compare the relative orientation between column density structures and magnetic field \citep{Soler_2013}. Applying this method to observations, a transition was found from parallel to perpendicular with increasing column density, indicating that the magnetic field plays an important role in the gas dynamics at cloud scales \citep{Planck_2016c,Soler_2017}.

A more detailed study attempted to connect the intensity structure of molecular lines from different gas tracers with the magnetic field in the Vela C Giant Molecular Cloud \citep{Fissel_2019ApJ...878..110F}. When comparing the results with different tracers, the authors found a transition from parallel to perpendicular as the critical density of the tracers increased, indicating a correlation between gas intensity structures and magnetic fields, which is consistent with the result of recent magnetohydrodynamics (MHD) simulations \citep{Mazzei}. However, \citet{Fissel_2019ApJ...878..110F} was only a case study, and their findings need to be validated in a larger sample.

The Orion Giant Molecular Cloud is the nearest high-mass star forming region \citep[d=414 pc,][]{Menten_2007}, and is well known for its hierarchical filamentary structure and strong star-forming activities \citep{Bally_2008}. The magnetic field properties in Orion A have been thoroughly investigated on both large \citep[e.g.,][]{Clark_1974, Houde_2004, Tahini_2022} and small scales \citep[e.g.,][]{Poidevin_2010,Pattle_2017,Soler_2019}. In the present work, we aim to examining 
the relationship between molecular gas structure and magnetic field at different scales. We make use of the spectral cube of three different molecular lines and then calculate different gas intensity structure maps to make a comparison with the magnetic field\footnote{We note that we compare the local relative orientation between intensity structures and magnetic field pixel by pixel. This is different from the velocity gradient technique (VGT), which requires an additional sub-block-averaging step \citep{Yuen_2017}. Due to this difference in methodology, our results cannot be directly compared to those obtained using VGT.}. Our goal is to decipher whether or not there is a similar trend in relative orientation with increasing column density within Orion A Giant Molecular Cloud, whether or not the correlation between gas structures and magnetic field changes when comparing them at different scales, and whether or not there are clear differences in relative orientation when using different gas tracers.

 We first introduce the Nobeyama-, SCUPOL-, and {\it{Planck}}-derived maps used in our analysis in Sect. \ref{observations}. Our methods of calculation and the ensuing results are described in Sect. \ref{sec:methods and results}. We show our main results in Sect. \ref{Results}.
 In Sect. \ref{sec:discussion}, we compare our results with the literature and discuss the implications of our work. A brief summary of our results is given in Sect. \ref{summary}.

\section{Observations} \label{observations}

\subsection{Planck dust polarization data}
\label{Planck_section}
For the analyses 
in this work, we calculated the magnetic field orientations from linearly polarized dust emission. {\it{Planck}} observations provide the linear polarization maps 
(Stokes Q and U) from 30 to 353 GHz in multiple frequency bands across the entire sky \citep{planck_2020a}. These maps serve as valuable tools for studying polarized emission from interstellar dust. On large scales, we adopt the publicly available PR3 data from the High-Frequency Instrument (HFI; \citealt{Lamarre_2010}) at 353 GHz \citep{Planck_2020c}. The maps of Stokes Q and U are initially at 5$^{\prime}$ spatial resolution ($\sim 0.6$ pc) and the pixel size of 1.71$^\prime$. We calculate the polarization position angle projected on the plane of sky from the Stokes parameters as  \begin{equation}
    \theta_\mathrm{p}^{\mathrm{g}}=\frac{1}{2} \arctan (U, Q).
    \label{equation 1}
\end{equation}
The magnetic field position angle can be obtained by adding 90$^{\circ}$ to the polarization angle: $\theta_\mathrm{B}^{\mathrm{g}}=\theta_\mathrm{p}^{\mathrm{g}}+\frac{\pi}{2}$. The downloaded {\it{Planck}} Q and U maps\footnote{We downloaded the data from http://pla.esac.esa.int/. If the data are downloaded from https://irsa.ipac.caltech.edu/applications/planck/, the derived polarization position angle should be $\theta_\mathrm{p}=\frac{1}{2} \arctan (-U, Q)$. } are in Galactic coordinates and the polarization position angle in equatorial coordinates can be derived using $\theta_\mathrm{p}^{\mathrm{e}}=\theta^{\mathrm{g}}_\mathrm{p}-\Delta \theta_{\mathrm{p}}^{\mathrm{g}-\mathrm{e}}$, where \begin{equation}
    \Delta \theta^{\mathrm{g-e}}_{\mathrm{p}}=\arctan [\frac{\cos \left(l-32.9^{\circ}\right)}{\cos b \cot 62.9^{\circ}-\sin b \sin \left(l-32.9^{\circ}\right)}]
\end{equation}
is the angle difference between Galactic and equatorial coordinates \citep{Corradi_1998}. The debiased polarization intensity $PI$ and the corresponding uncertainties $\sigma_{PI}$ can be derived using the modified asymptotic estimator (pMAS; \citealt{Plaszczynski_2014}). The rms noise level of the stokes Q and U map is similar, with a value of 0.26 $\mathrm{mK_{CMB}}$ (hereafter $\sigma_{QU})$. In our analysis, we only use the polarization data that have a sufficient signal-to-noise ratio ($PI/\sigma_{PI}>3$). The uncertainties of polarization position angle can be calculated using $\sigma_{\theta}\sim\frac{\sigma_{QU}}{2} \sqrt{{1}/{(Q^{2}+U^{2})}}$.

\subsection{JCMT dust polarization data}

On small scales, we used data from the James Clerk Maxwell Telescope (JCMT)/SCUPOL catalog \citep{Matthews_2009}  to measure the relative orientation of the magnetic
field. This catalog is a combination of calibrated and reduced data observed between 1997 and 2005 at 850 $\mu$m by the polarimeter of the Submillimeter Common-User Bolometer Array (SCUBA) on the James Clerk Maxwell Telescope. There are two regions  of Orion A for which high-resolution polarization data are available: OMC-1 and OMC-2/3. The individual observations were combined to create composite Stokes I, Q, and U maps. The data are sampled on a 10$^{\prime \prime}$ per pixel grid in J2000 coordinates and the effective beam size of the map is 20$^{\prime \prime}$ ($\sim 0.04$ pc). We calculate the polarization position angle projected on the plane of sky from the Stokes parameters as \begin{equation}
    \theta_\mathrm{p}^{\mathrm{e}}=\frac{1}{2} \arctan (U, Q)
    \label{equation 2}
.\end{equation}
The derived position angle is in equatorial coordinates, and there is no need for transformation. We applied the same method mentioned in Sect. \ref{Planck_section} to calculate the uncertainties, and the rms noise of stokes Q and U maps is approximately 0.1 mJy/beam within OMC-1 and 0.03 mJy/beam within OMC-2/3.

\subsection{Nobeyama molecular line data}

To study the gas structures in Orion A, we selected data from the NRO Star Formation Legacy Project\footnote{{{https://jvo.nao.ac.jp/portal/nobeyama/}}} based on observations at the Nobeyama Radio Observatory (NRO). The observations cover $^{12}\mathrm{CO}$ ($J=1-0$), $^{13}\mathrm{CO}$ ($J=1-0$), $\mathrm{C}^{18}\mathrm{O}$ ($J=1-0$), $\mathrm{N}_2\mathrm{H}^{+}$ ($J=1-0$), and CCS ($J_N=8_7-6_7$) lines toward some nearby star-forming regions (M17, Orion A, and Aquila Rift). Detailed information regarding the observations is summarized in \citet{Nobeyama_star_forming_project_overview}.  In this paper, we use three different molecular lines with significant extended emission: $^{12}\mathrm{CO}$ (J=1-0), $^{13}\mathrm{CO}$ (J=1-0), and $\mathrm{C}^{18}\mathrm{O}$ (J=1-0). The $^{12}\mathrm{CO}$, $^{13}\mathrm{CO}$, $\mathrm{C}^{18}\mathrm{O}$ data are convolved to 21.7$^{\prime \prime}$ beam size and reprojected to a common 7.5$^{\prime \prime}$$\times$7.5$^{\prime \prime}$ grid. The velocity resolution is $0.1 \mathrm{~km} \mathrm{~s}^{-1}$ and the noise levels of $^{12}\mathrm{CO}$, $^{13}\mathrm{CO}$, and $\mathrm{C}^{18}\mathrm{O}$ lines are at a range of $0.5-1.5$ K, $0.2-0.3$ K, and $0.26-0.3$ K, respectively. For the sake of comparison with {\it{Planck}} observations, we use the maps smoothed to an angular resolution of 5$^{\prime}$, and we directly compare the maps based on SCUPOL and Nobeyama observations on small scales without smoothing given their similar angular resolution.

\section{Methods} \label{sec:methods and results}

\subsection{Detailed procedure of comparison}

\begin{figure*}[!bt]
        \centering
        \includegraphics[width=1.0\linewidth]{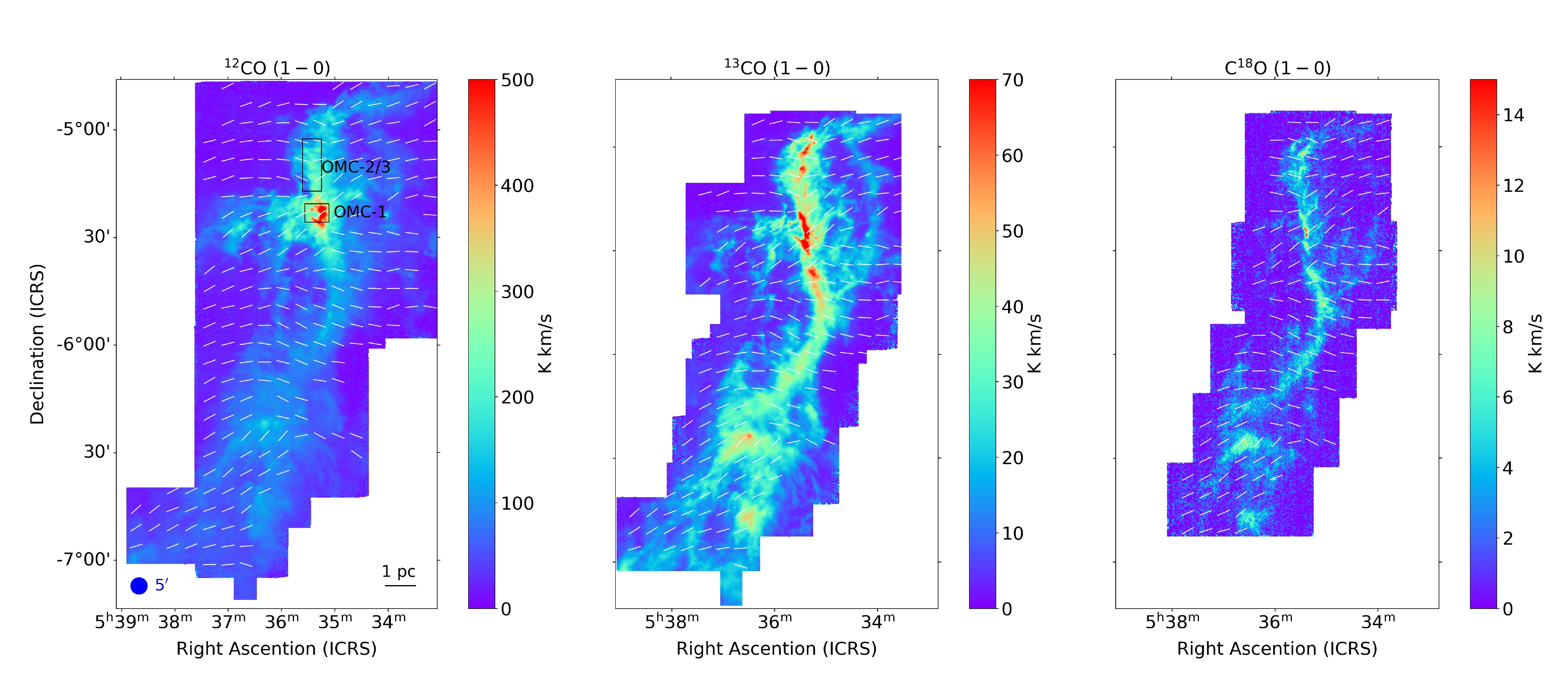}
        \caption{Nobeyama moment-0 maps for $^{12}$CO, $^{13}$CO, and C$^{18}$O with the resolution of 21.7$^{\prime\prime}$. The white vectors represent the orientation of {\it{Planck}} magnetic fields. Only segments with sufficient signal-to-noise ratio (PI/$\sigma_{\mathrm{PI}}>$3) are plotted. The blue circle in the bottom left marks the resolution of {\it{Planck}} magnetic fields. The black rectangles represent the specific regions with high-resolution polarization observations.}
    \label{m0_all}
\end{figure*}

\begin{figure}[!bt]
\centering
\includegraphics[width=1.0\linewidth]{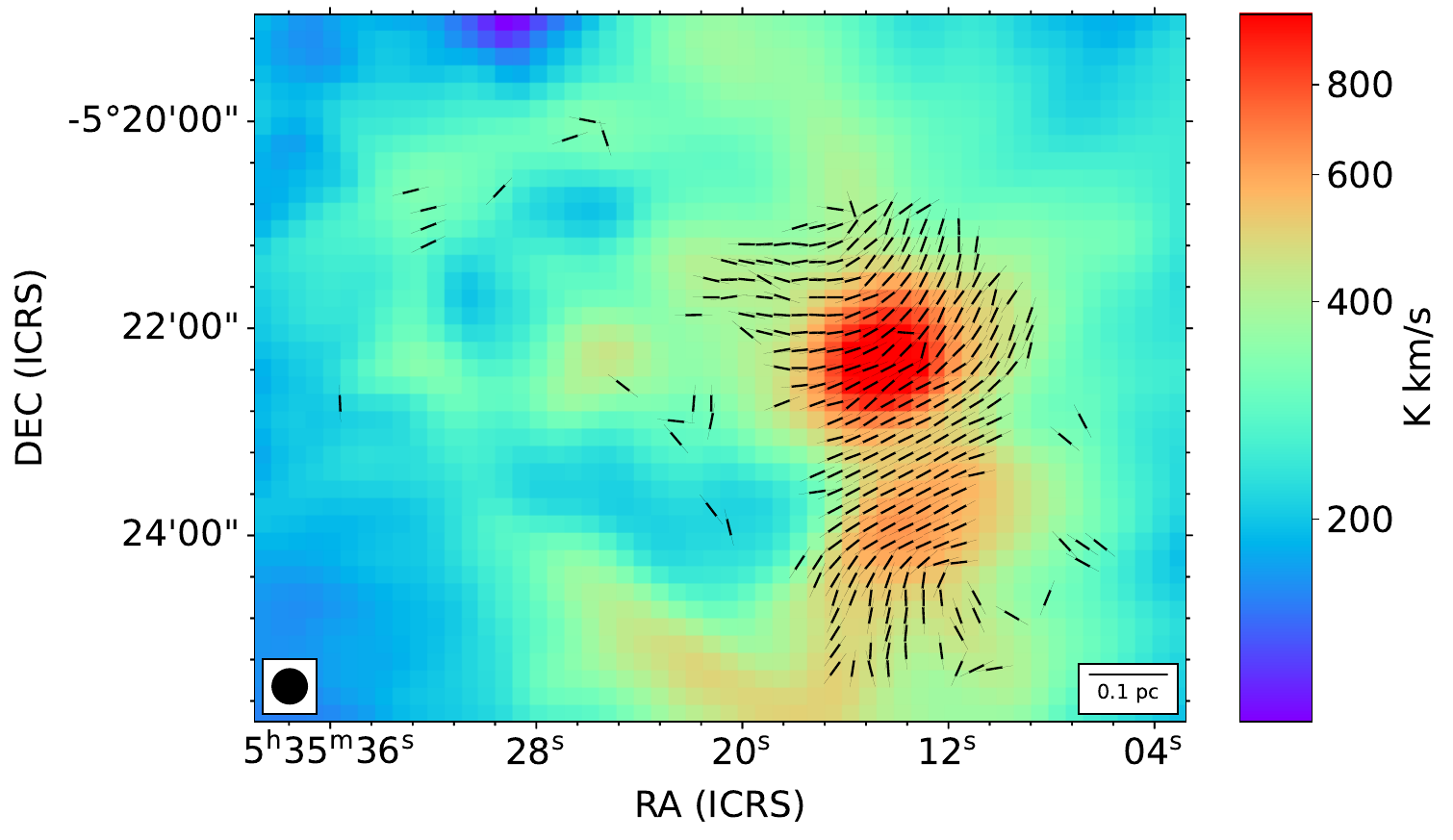}
\caption{Magnetic field vectors overlaid on the $^{12}$CO moment-0 map of OMC-1. Only the segments with stokes I signal-to-noise ratio (S/N$>25$) and PI/$\sigma_{\mathrm{PI}}>$3 are plotted. The black circle in the bottom-left corner marks the resolution of the SCUPOL magnetic fields. }
\label{jcmt_omc1}
\end{figure}

\begin{figure}[!bt]
\centering
\includegraphics[width=1.0\linewidth]{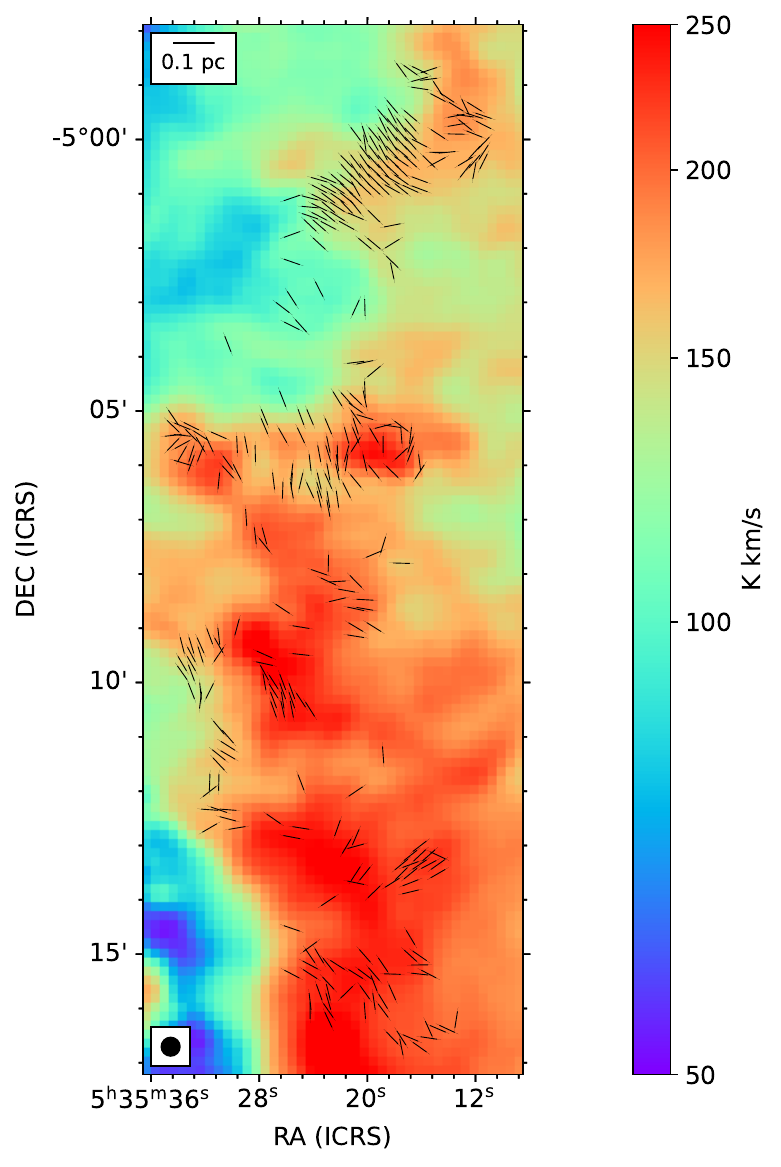}
\caption{Magnetic field vectors overlaid on the $^{12}$CO moment-0 map of OMC-2/3. Only the segments with stokes I signal-to-noise ratio (S/N$>25$) and PI/$\sigma_{\mathrm{PI}}>$3 are plotted. The black circle in the bottom-left corner marks the resolution of the SCUPOL magnetic fields. }
\label{jcmt_omc23}
\end{figure}

Here we describe the procedure we used to compare maps. On large scales, we initially smoothed the spectral line data cube to match the angular resolution of {\it{Planck}} observations and then compared the relative orientation of gas-intensity gradients with the polarization position angle (perpendicular to the magnetic field) at low angular resolution. In special regions with high-resolution SCUPOL polarization observations, we directly calculated the gradient vector field of each map and compared it with the orientation of polarization. 

\subsection{Column-density maps} \label{column density}
\par The moment-0 map was created by integrating the emission between 0 and $18 \mathrm{~km} \mathrm{~s}^{-1}$. To exclude the impacts of unreliable pixels, we selected pixels whose spectra exhibit at least five velocity channels with a brightness temperature of higher than 8$\sigma$ for $^{12}\mathrm{CO}$, 5$\sigma$ for $^{13}\mathrm{CO}$, and 3$\sigma$ for $\mathrm{C}^{18}\mathrm{O}$. The derived moment-0 maps for the Orion A region are shown in Fig. \ref{m0_all}, and zoomed-in  $^{12}\mathrm{CO}$ moment-0 maps for OMC-1 and OMC-2/3 with overlaid JCMT-derived magnetic field vectors are shown in Figs. \ref{jcmt_omc1} and \ref{jcmt_omc23}.

Previous studies show that the relative orientation between gas structures and magnetic field changes from parallel to perpendicular with increasing column density \citep[e.g.,][]{Planck_2016c,Soler_2017}. 
In our study, we estimated the column-density structures using the $^{12}\mathrm{CO}$ and $^{13}\mathrm{CO}$ lines \citep[see detailed derivation from][and references therein]{Mangum&shirley_2015,Li_2018}. Firstly, assuming local thermodynamic equilibrium (LTE) and an always optically thick $^{12}\mathrm{CO}$ line, we calculated the excitation temperature of $^{12}\mathrm{CO}$ according to the following formula:\begin{equation}
    T_{\text{ex}}=5.532\left[\ln \left(1+\frac{5.532}{T_{\text {peak}}+0.819}\right)\right]^{-1}
,\end{equation}
where $T_{\text{peak}}$ is the maximum brightness temperature of the $^{12}\mathrm{CO}$ line. The optical depth of $^{13}\mathrm{CO}$ can then be calculated as \begin{equation}
    \tau^{13}(v)=-\ln \left[1-\frac{T_{\mathrm{mb}}(^{13}\mathrm{CO})}{5.29}\left(\left[e^{5.29 / T_{\mathrm{ex}}}-1\right]^{-1}-0.164\right)^{-1}\right]
,\end{equation}
where $T_{\mathrm{mb}}(^{13}\mathrm{CO})$ is the brightness temperature of the $^{13}\mathrm{CO}$ line. Assuming that the emission of $^{13}\mathrm{CO}$ is optically thin, we can use the approximate relation \citep{Pineda_2010}: 
 \begin{equation}
    T_{\mathrm{ex}} \int \tau d v \approx \frac{\tau_0}{1-e^{-\tau_0}} \int T_{\mathrm{mb}} d v
,\end{equation}
where $\tau_0$ is the peak optical depth of the $^{13}\mathrm{CO}$ emission line. Assuming $ \mathrm{H_2} /{ }^{13} \mathrm{CO}=7.1\times 10^5$ \citep{Frerking_1982ApJ...262..590F}, we can obtain the total column density of molecular hydrogen: \begin{equation}
     N(\mathrm{H}_{2})^{^{13} \mathrm{CO}}= 1.72 \times 10^{20} \frac{\tau_0}{1-e^{-\tau_0}} \\ \times \frac{1+0.88 / T_{\mathrm{ex}}}{1-e^{-5.29 / T_{\mathrm{ex}}}} \int T_{\mathrm{mb}}{(^{13} \mathrm{CO})}  dv 
.\end{equation}

One thing we must note is that the $^{13}\mathrm{CO}$ line might be optically thick in some dense regions within Orion A, but the derived column density maps were only used to examine the relation between the orientation and column density, and so this is unlikely to significantly affect our results.

\subsection{Calculation of relative orientations}\label{sec:orientation}

We determined the relative angle between the intensity structures and the magnetic field using a method similar to that described in \citet{Soler_2013}.  The angle $\phi$, between the tangent to the local intensity structure contours and the magnetic field can be calculated using the following formula \citep{Fissel_2019ApJ...878..110F}: \begin{equation}
\phi=\arctan \ (|\nabla I \times \theta_{\mathrm{p}}|, \nabla I \cdot \theta_{\mathrm{p}})
,\end{equation}
where $\theta_{\mathrm{p}}$ marks the polarization position angle perpendicular to the orientation of the magnetic fields, $\nabla I$ represents the orientation of the intensity gradient that is perpendicular to the contours of intensity, the relative orientation angle $\phi$ is within the range $[0^{\circ}, 90^{\circ}],$ and the angle outside the range can be transformed into a uniquely determined angle within the limit.

The gradient technique was employed to characterize the orientation of structures in a scalar field. Within a 3$\times$3 pixel grid centered around each pixel, the gradient of gas intensity structures was derived using a similar method to that described in \citet{Sokolov_2019}. We used a first-degree bivariate polynomial function to describe the gradient field: $f(\alpha, \delta)=I_0 +a \Delta \alpha+b \Delta \delta$, where $I_0$ represents the intensity of the central point, and $\Delta \alpha$ and $\Delta \delta$ represent the pixel offsets. To ensure precision, only pixels with sufficient signal-to-noise ratio were included in the gradient analysis. In addition, we selected pixels with at least eight valid data points in the surrounding 3$\times$3 pixel grid centered on those pixels. A least-squares method was employed to derive the gradient field toward each pixel:\begin{equation}
    p({r})={\mathop{argmin}\limits_{a, b}{\sum_{\Delta \alpha,\Delta \delta\le1} {(I({r}^{\prime})-f({r}^{\prime}, a, b))^{2}}}}
.\end{equation}
The orientation of gradient and its uncertainty can be given by $\nabla I= \arctan \ (b, a)$ and $\sigma_{\nabla I}=\frac{1}{a^2+b^2}\sqrt{a^2\sigma_b^2+b^2\sigma_a^2}$ \citep{Planck_2016c}.

We assessed the relative orientation of the magnetic field in relation to the gas structures using the HRO technique \citep{Soler_2013}. The map was segmented into seven intensity bins, each containing an equal number of selected pixels based on their intensity values. We used nine angle bins, each with a width of $10^{\circ}$. In addition, we replaced the intensity structures with column density to examine whether there is a transition from parallel to perpendicular alignment with increasing column density. Uncertainties in the HRO plot were estimated through a Monte Carlo method by adding the uncertainties of each parameter pixel by pixel. We repeated this method 10000 times, and took the mean value and standard deviation as the result and the uncertainty, respectively.

\subsection{Statistical study of relative orientation}
\subsubsection{Histogram shape parameter $\xi$}
We analyzed the changes in the HRO shape with column density, employing the histogram shape parameter \citep{Planck_2016c}:\begin{equation}
    \xi=\frac{A_{\mathrm{c}}-A_{\mathrm{e}}}{A_{\mathrm{c}}+A_{\mathrm{e}}}
.\end{equation}
Given that the angle $\phi$ ranges from 0 to 90 degrees in our analysis, $A_\mathrm{c}$ represents the area concentrated around 0 degrees in the histogram ($0^{\circ}<\phi<20^{\circ}$) and $A_\mathrm{e}$ is the area concentrated around 90 degrees in the histogram ($70^{\circ}<\phi<90^{\circ}$). A positive $\xi$ ($>$ 0) indicates a parallel alignment between intensity structure contours and the magnetic field, while a negative $\xi$ ($<$ 0) suggests that the intensity structure is more likely to be perpendicular to $\theta_\mathrm{B}$. We use a linear function to investigate the correlation between $\xi$ and the total gas column density for each tracer:\begin{equation} \label{fitting_equation}
    \xi=C_{\mathrm{HRO}}[ \ \log _{10}(N_{\mathrm{H}} / \mathrm{{cm}^{-2}})-X_{\mathrm{HRO}}]
,\end{equation}
where $C_{\mathrm{HRO}}$ is the slope of the linear relation, and $N_{\mathrm{H}}$ is the total gas column density. The molecular gas column density $N_{\mathrm{H_2}}$ can be transferred to the $N_{\mathrm{H}}$ through the relation $N_{\mathrm{H}}=N_{\mathrm{HI}}+2 \times N_{\mathrm{H_2}}$. However, because we only focus on regions with high column density, the contribution of $N_{\mathrm{HI}}$ can be neglected \citep{Sternberg_2014}. We therefore derived the total gas column density using $N_{\mathrm{H}}=2 \times N_{\mathrm{H_2}}$. The value of $X_{\mathrm{HRO}}$ for the $\log _{10}(N_{\mathrm{H}} / \mathrm{{cm}^{-2}})$ represents the transition point where the relative orientation changes from parallel to perpendicular. The Pearson correlation coefficient and p-value are computed to measure the significance of the linear correlation. The histogram shape parameter is used to measure the preference of the angle groups for pure parallel or perpendicular alignments.

\subsubsection{Alignment measure parameter\label{Section 3.3.3}}

\par \citet{Jow_2018} introduced the projected Rayleigh statistic (PRS) as a test for nonuniform relative orientation between two pseudo-vector fields, and it has since been used by several authors \citep[e.g.,][]{Soler_2019,Beuther_2020}. However, when comparing the results of different gas tracers, direct comparison of PRS values is not feasible because it is not a normalized parameter and can be influenced by the number of data points. Here, we use the normalized alignment measure (AM) parameter, as described by \citet{Gonz_lez_Casanova_2017} and \citet{Lazarian_2018}:\begin{equation}\label{fitting_equation_AM}
    \mathrm{AM}=2\left\langle\cos^{2} (\phi)\right\rangle-1
.\end{equation}

If the results of AM $ \rightarrow 1$, this indicates a significant parallel alignment. Alternatively, AM $ \rightarrow -1$  indicates a significant perpendicular alignment. In cases where there is no preference for angle distribution, the value of AM will be close to 0. We use the same method as in Eq. \ref{fitting_equation} to test the linear correlation between AM and column density. The physical meaning of the AM parameter is similar to the physical meaning of the histogram shape parameter. The difference lies in $\xi$ being employed to measure the preference of angle groups for purely parallel or perpendicular alignments, while AM is used to characterize global orientation distributions. If we find similar trends in both $\xi$ and AM, the alignment preference will be more reliable.

\begin{figure*}[ht!]
    \centering
    \begin{subfigure}[a]{1.0\textwidth}
        \centering
        \includegraphics[width=0.99 \linewidth]{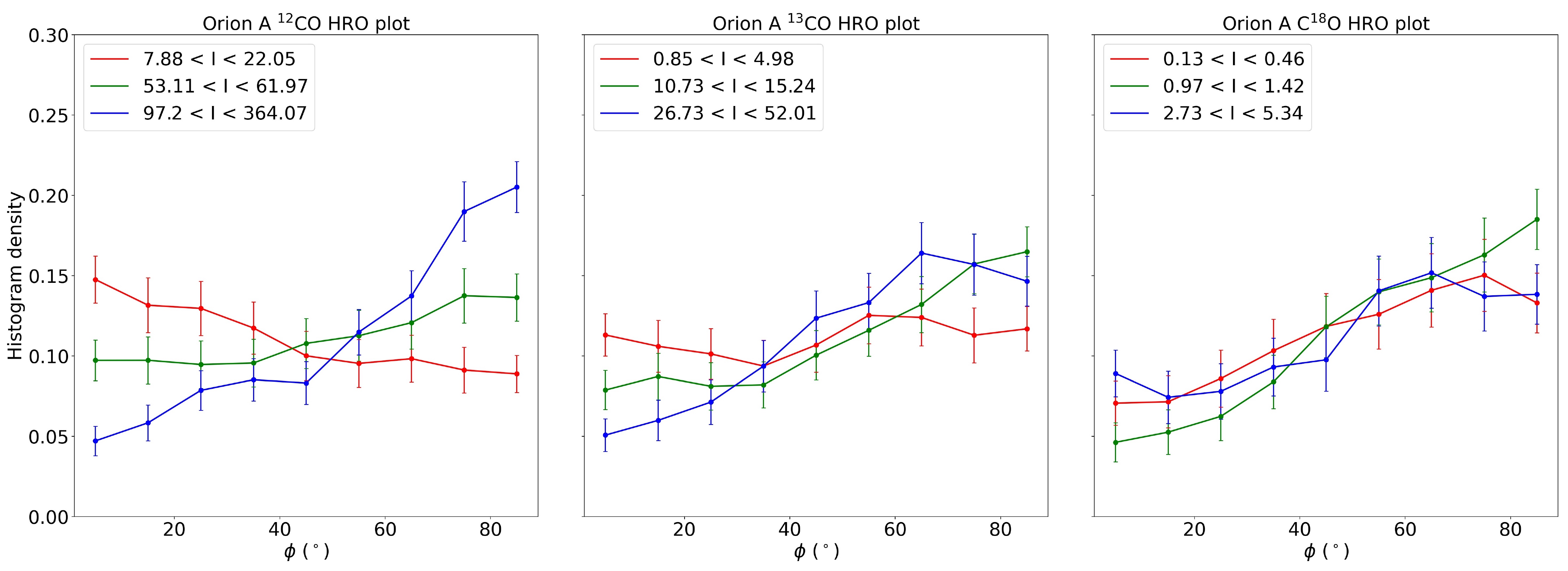}
        \caption{}
        \label{fig:OrionA_HRO}
    \end{subfigure}
    \begin{subfigure}[b]{1.0\textwidth}
        \includegraphics[width=\linewidth]{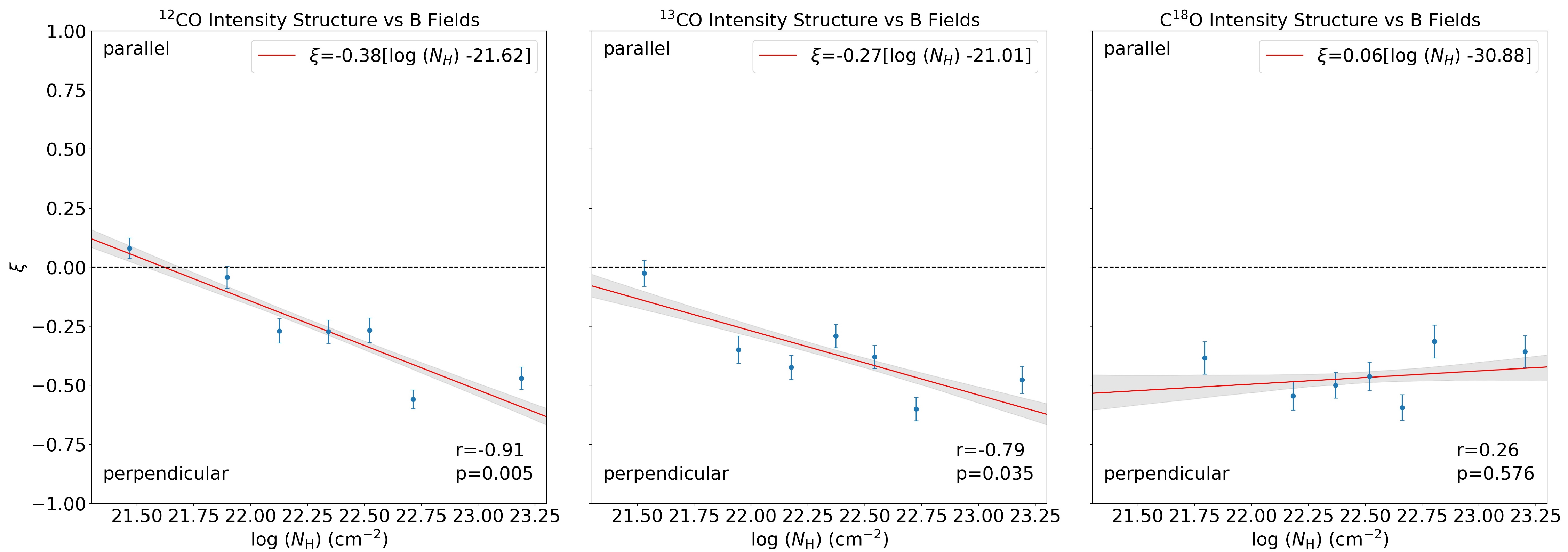}
        \caption{}
        \label{fig:OrionA_xi}
    \end{subfigure}
    \begin{subfigure}[c]{1.0\textwidth}
        \includegraphics[width=\linewidth]{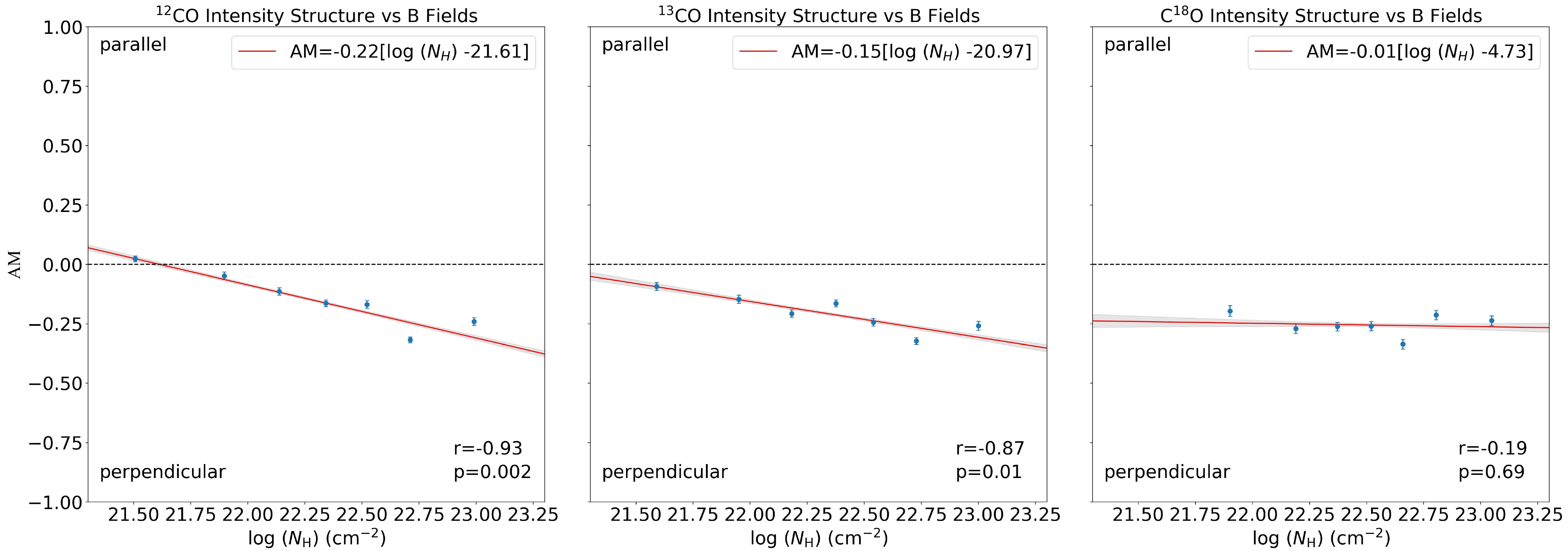}
        \caption{}
        \label{fig:OrionA_AM}
    \end{subfigure}
    \caption{ (\textbf{a}): Histogram of relative orientations between intensity structures for different tracers and magnetic field in the whole Orion A region. The red, green, and blue solid lines represent the angle distribution in the lowest, medium, and highest intensity bins, respectively.  (\textbf{b}): Histogram shape parameter calculated in different column-density bins in the whole Orion A region. The red solid line is the optimal linear fitting results between $\xi$ and column density. The gray area represents the 1$\sigma$ confidence interval of the fitted line. (\textbf{c}): Same as (\textbf{b}), but for the relation between AM and column density.}
    \label{fig:OrionA}
\end{figure*}

\section{Results}
\label{Results}
\subsection{Results on large scales}

We examine the relative orientation between gas intensity structures and the magnetic field across the entire Orion A region at a low resolution ($\sim$ 0.6 pc).  The HROs in different intensity bins are presented in Fig. \ref{fig:OrionA}(a). The figure illustrates a clear difference in angle distribution among various intensity bins for $^{12}$CO, ranging from a weak preference for parallel alignment to a preference for perpendicular alignment. This difference becomes less clear with decreasing optical depth of gas tracers. Both $\xi$ and AM exhibit a clear anti-correlation with column densities for $^{12}$CO and $^{13}$CO. Regarding C$^{18}$O, there appears to be a preference for the orientation between intensity structures and magnetic field to be perpendicular across all column density bins, and there is no correlation between $\xi$/AM and column density. The transition from parallel to perpendicular is consistent with some previous works \citep{Planck_2016c, Soler_2019} and MHD simulations \citep{Soler_2013, Chen_2016}, indicating a trans-to-sub-Alfvénic state (see \citealt{Liu_2022} for a review). The variation in results from different gas tracers can be explained by the fact that different gas molecules trace different density regions. In high-density regions, the orientation between intensity structures and the magnetic field is more likely to be perpendicular. Our results are consistent with the deduction described in \citet{Fissel_2019ApJ...878..110F}.

\subsection{Results on small scales}

\begin{figure*}[ht!]
    \centering
    \begin{subfigure}[a]{1.0\textwidth}
        \centering
        \includegraphics[width=0.99 \linewidth]{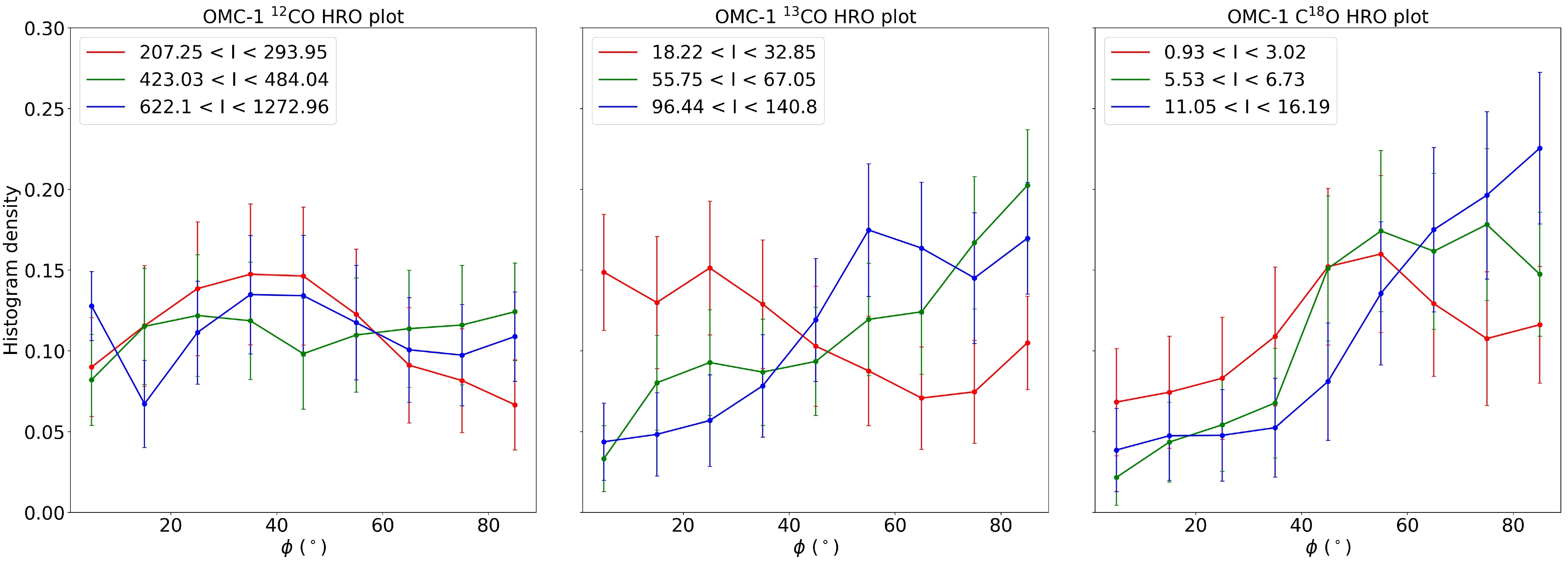}
        \caption{}
    \end{subfigure}
    \begin{subfigure}[b]{1.0\textwidth}
        \includegraphics[width=\linewidth]{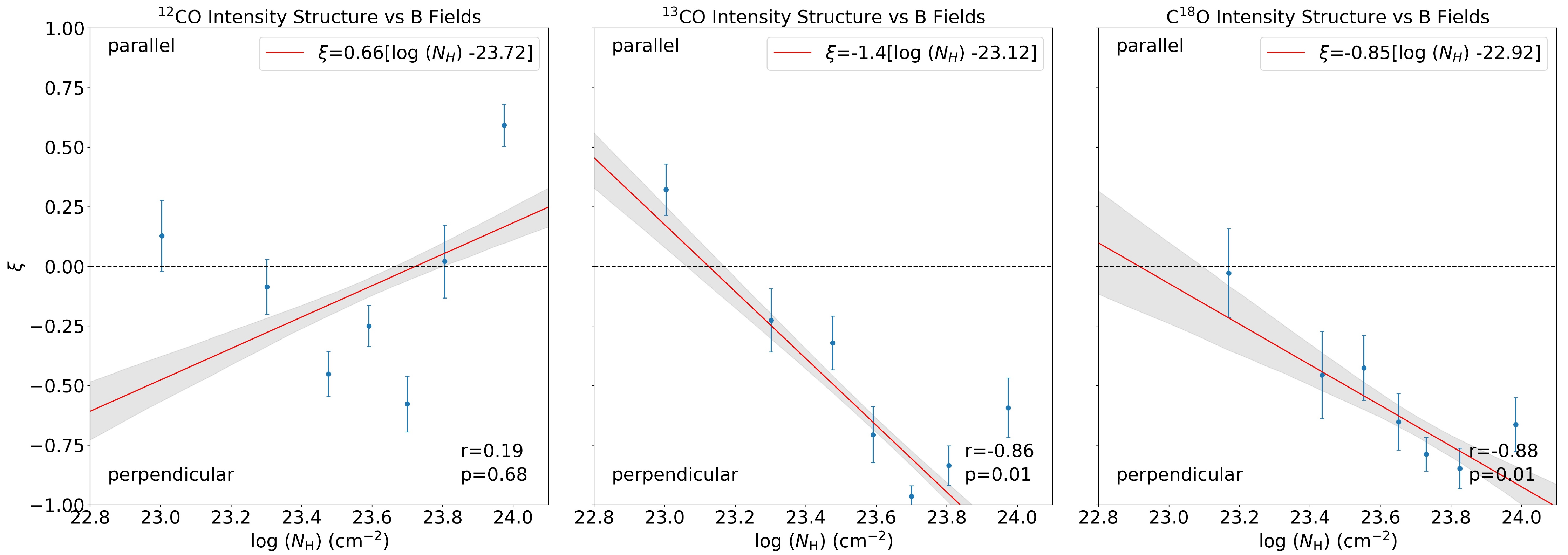}
        \caption{}
    \end{subfigure}
    \begin{subfigure}[c]{1.0\textwidth}
        \includegraphics[width=\linewidth]{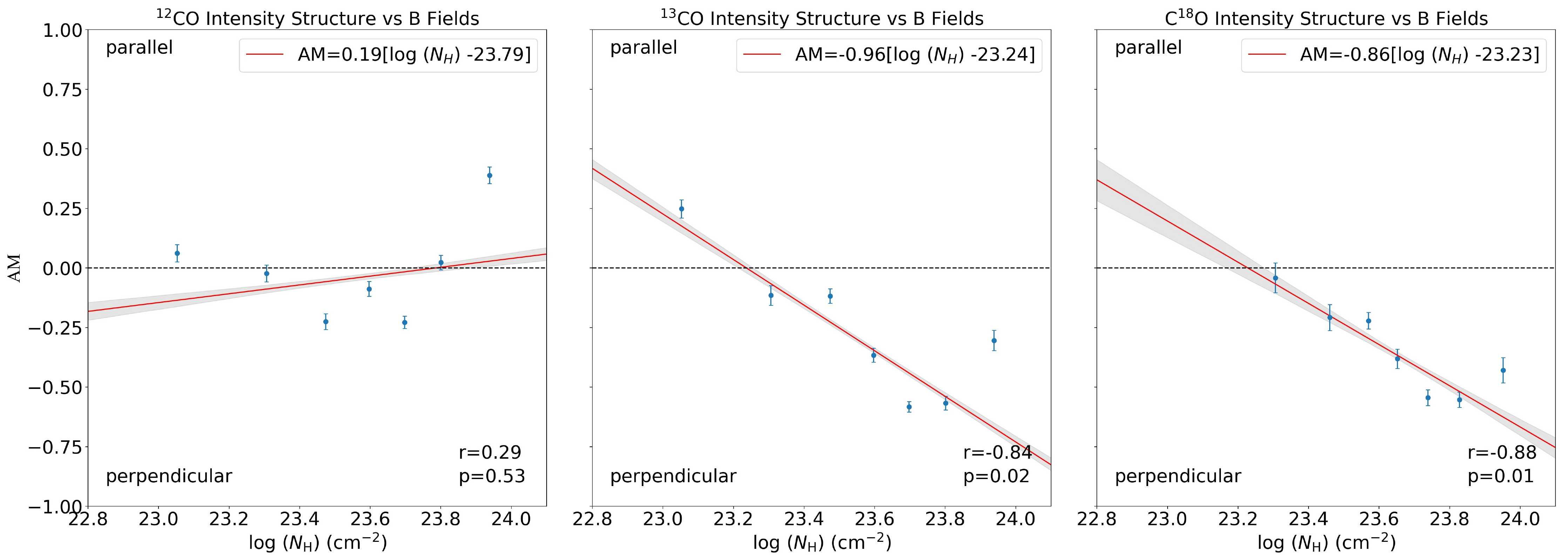}
        \caption{}
    \end{subfigure}
    \caption{Same as Fig. \ref{fig:OrionA}, but for the relative orientation in OMC-1.}
    \label{fig:OMC1}
\end{figure*}

We compare the relative orientation between gas intensity structures and magnetic field in two specific areas (OMC-1, OMC-2/3) within the Orion A region, where high-resolution polarization observations are available ($\sim$ 0.04 pc). There are significantly higher column densities traced in these areas. In this section, we present the respective results for each region.

\subsubsection{Results in OMC-1}

Figure \ref{fig:OMC1}(a) illustrates the HRO plots within various intensity bins, while Figs. \ref{fig:OMC1}(b) and \ref{fig:OMC1}(c) display the correlation between the relative orientation and column densities. In contrast to the findings on large scales, the results in the OMC-1 region differ. We find that the relative orientation between gas intensity structures and magnetic field shows no correlation with column density in the $^{12}\mathrm{CO}$ line, but exhibits a clear transition from parallel to perpendicular alignment with increasing column density in the $^{13}\mathrm{CO}$ and $\mathrm{C}^{18}\mathrm{O}$ lines. Despite slightly increased uncertainties in calculations due to the limited number of selected pixels, the linear relationship maintains statistical significance after performing Monte Carlo simulations as outlined in Sect. \ref{sec:orientation} ($|r|>0.7, p<0.05$, and the $1\sigma$ confidence interval of the best-fit line shows relatively little spread). This phenomenon can be interpreted by the fact that the $^{12}\mathrm{CO}$ line is a low-density gas tracer. In the OMC-1 region, $^{12}\mathrm{CO}$ is optically thick, and its intensity structure cannot effectively trace the density structure but only the surface of the cloud. On the other hand, the denser gas tracers, $^{13}\mathrm{CO}$ and $\mathrm{C}^{18}\mathrm{O}$, prove to be more suitable for tracing the density structures.

\subsubsection{Results in OMC-2/3} 

\begin{figure*}[ht!]
    \centering
    \begin{subfigure}[a]{1.0\textwidth}
        \centering
        \includegraphics[width=0.99\linewidth]{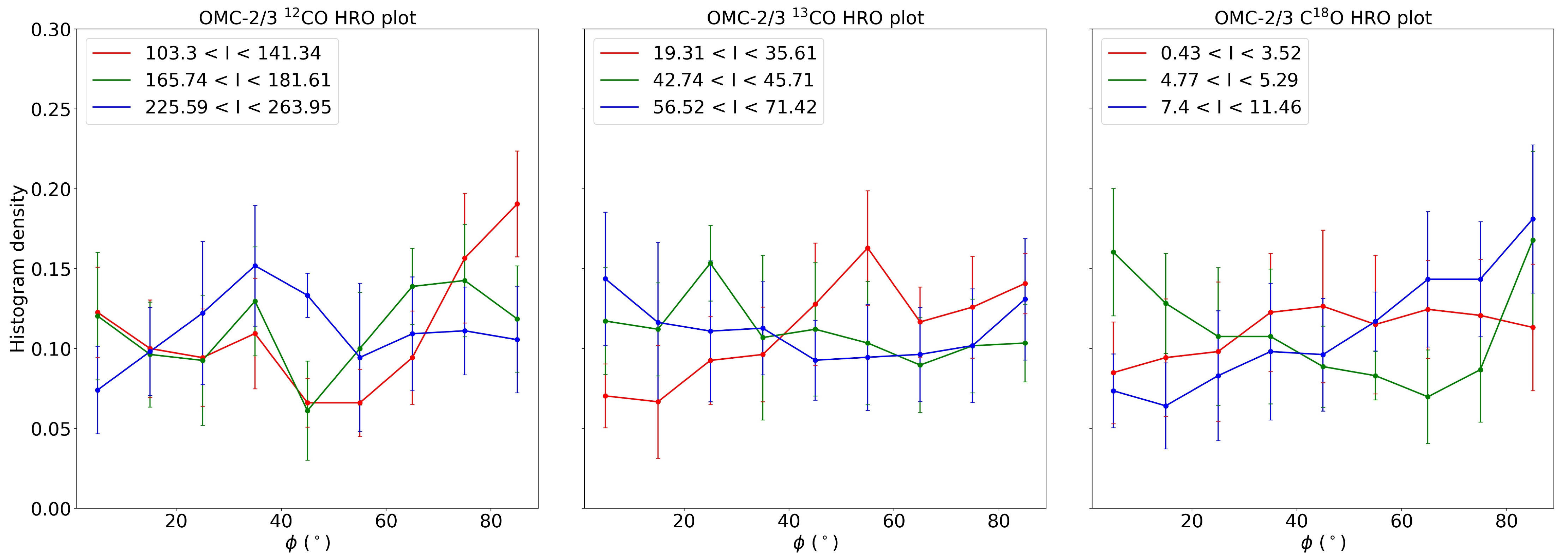}
        \caption{}
    \end{subfigure}
    \begin{subfigure}[b]{1.0\textwidth}
        \centering
        \includegraphics[width=\linewidth]{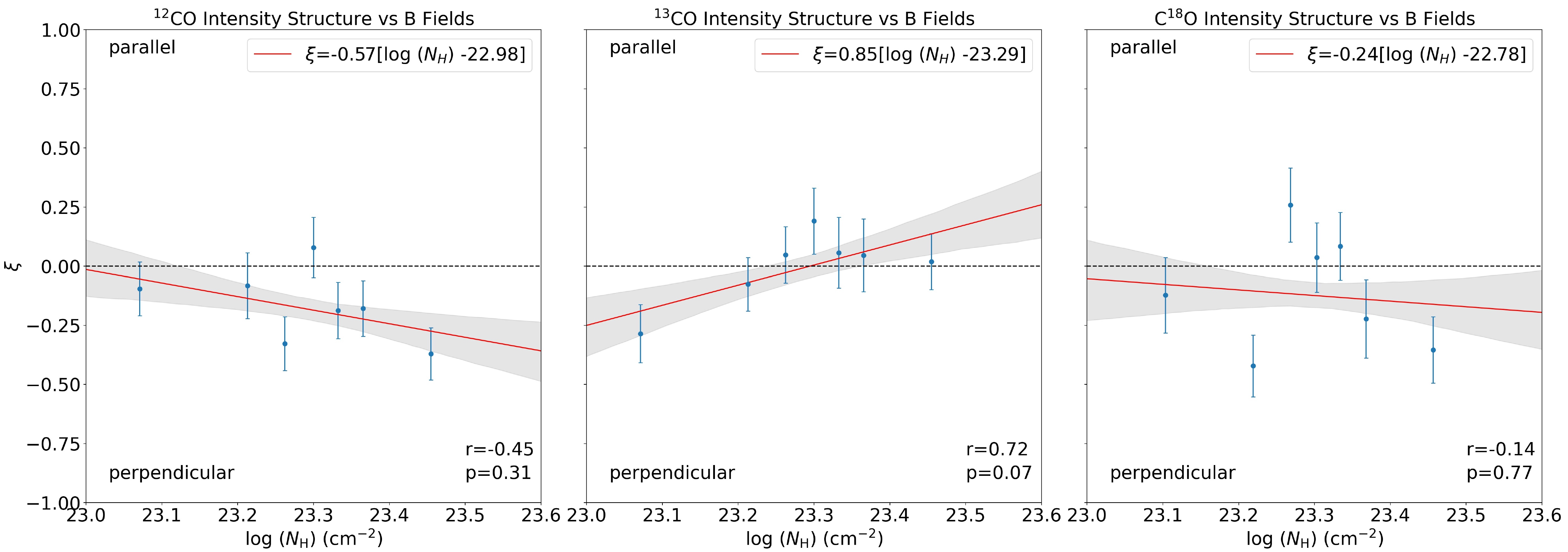}
        \caption{}
    \end{subfigure}
    \begin{subfigure}[c]{1.0\textwidth}
        \includegraphics[width=\linewidth]{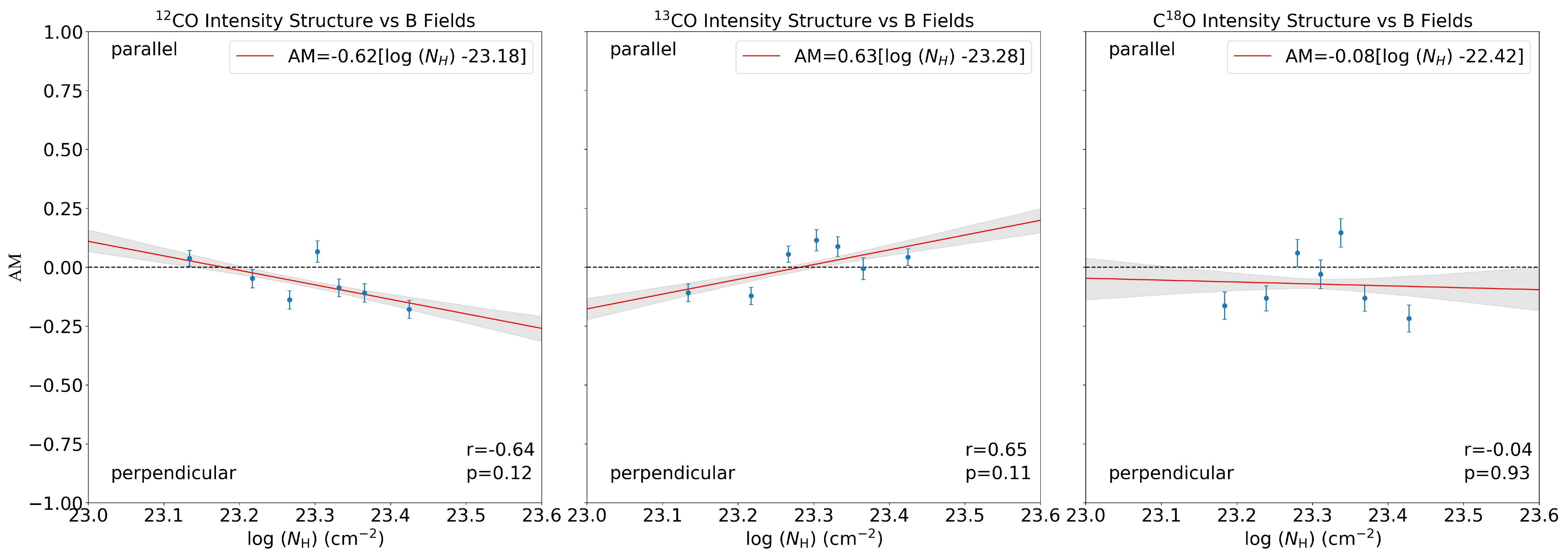}
        \caption{}
    \end{subfigure}
    \caption{Same as Fig. \ref{fig:OrionA}, but for the relative orientation in OMC-2/3.}
    \label{fig:OMC2/3}
\end{figure*}

Figure \ref{fig:OMC2/3}(a) presents the HRO plots within various intensity bins, while Figs. \ref{fig:OMC2/3}(b) and \ref{fig:OMC2/3}(c) show the correlation between relative orientation and column densities now for the OMC-2/3 regions. From the figure, it is evident that there is no statistically significant trend in orientation distribution with increasing column density (all p-values are greater than 0.05). Both $\xi$ and AM are close to zero across all column-density bins, indicating that the relative orientation between intensity structures and the magnetic field is more likely to be randomly distributed in all tracers. We conclude that there is no systematic trend in angle distribution with increasing column density in the OMC-2/3 region.

A possible explanation for this result is that there are numerous star-forming regions along the long filaments \citep{Chini_1997} and they are at different evolutionary stages \citep{Castets_1995}. The relative alignment between gas structures and the magnetic field is complex in star-forming regions at small scales. The impact of feedback from star-forming activity can lead to the realignment of structures, such as the reversal of the orientation to parallel via gas flows \citep{Pillai_2020}, or the induction of orientation through gravity-dragged rotation \citep{Beuther_2020,Sanhueza_2021}. The relative alignment between gas structures and magnetic field in high-mass star-forming regions should be investigated case by case, depending on scale, the tracers used, and the evolutionary stages of the structures investigated. The results in OMC-2/3 are discussed further in Sect. \ref{sec:projection effects}.

\section{Discussion} \label{sec:discussion}

\begin{figure*}[!bt]
        \centering
        \includegraphics[width=1.0\linewidth]{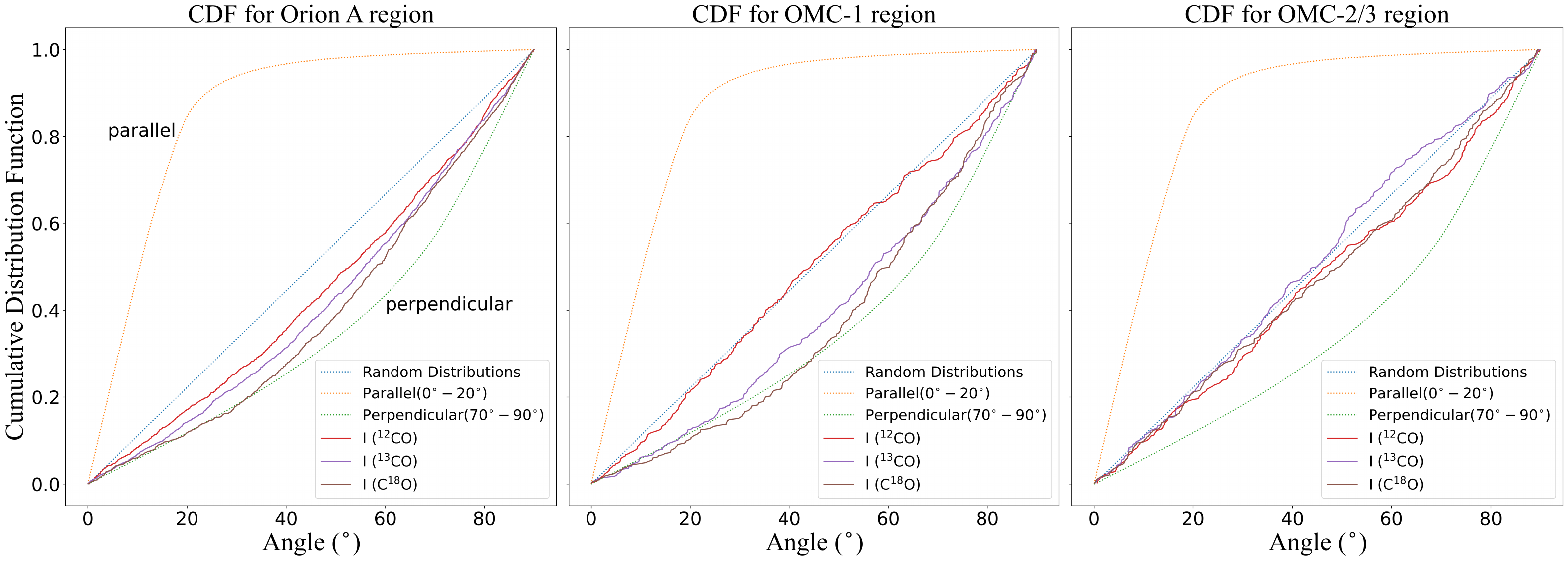}
        \caption{Cumulative distribution functions of relative orientation between intensity structures and magnetic field vectors and the projected $\theta_{3D}$ for Orion A, OMC-1, and OMC-2/3. }
    \label{CDF}
\end{figure*}

\subsection{Projection effects}
\label{sec:projection effects}

We measured the relative orientation between intensity structures and magnetic field on the plane of the sky. Due to projection effects, the angle distribution in 2D is not the same as the angle distribution in 3D space \citep{Seifried_2020}. To examine the influence of projection effects on the measured angle distribution, we used a Monte Carol simulation method employed in a previous study \citep[see][]{Stephens_2017}. Here we briefly introduce the method. First, we created $10^6$ pairs of two random unit vectors within a unit sphere and calculated the angle between the two vectors in 3D space. The angle is within the limit $[0^{\circ}, 90^{\circ}]$. We then projected the angle from 3D to 2D space and plotted the cumulative distribution functions (CDFs) in 2D in three angle bins: (1) purely parallel: angle range from $0^{\circ}$ to $20^{\circ}$ in 3D space, (2) random distribution: angle range from $0^{\circ}$ to $90^{\circ}$ in 3D space, and (3) purely perpendicular: angle range from $70^{\circ}$ to $90^{\circ}$ in 3D space. Finally, we plotted the CDF for our data and compared it with the projected 3D angle.

We present the CDF of $\phi$ and the projected $\theta_{\mathrm{3D}}$ in Fig. \ref{CDF}. In the OMC-2/3 region, a random angle distribution is observed for all tracers. However, in Orion A at a low resolution and in the OMC-1 region at a high resolution, a clear transition from random to perpendicular orientation is evident when changing from low-density tracers to high-density tracers, indicating that magnetic fields are dynamically important at both scales. Although the results are influenced by the integration effect along the line of sight, the close-to-perpendicular alignment of the two projected vectors in 2D must also indicate perpendicular alignment in 3D, which is in line with the conclusions of previous studies \citep{planck_HRO_diffuse,Planck_2016c}. 

The perpendicular alignment can be explained by the contraction scenario of a magnetized, self-gravitating, static collapsing cloud \citep{Mouschovias_1976a,Mouschovias_1976b}. On small scales, given that the spatial resolution of polarization observations is comparable to the typical width of filaments \citep{Arzoumanian_2019, Andre_2022}, the filament formation process may affect the magnetic field structures. Earlier studies discussed the relationship between the orientation of the magnetic field and that of the filaments, but no consistent alignment preference was found \citep{Arzoumanian_2021, Baug_2021}. In our work, perpendicular alignment is only observed in the OMC-1 region. One possible interpretation is that the magnetic field orientation is almost perpendicular to the major axis of the filament in OMC-1 but that there is no similar alignment in OMC-2/3. The gravitational interaction inside the filaments and the BN/KL outflow helps to shape the magnetic field geometry in OMC-1, suggesting feedback from star-forming activity may play an important role in the evolution of magnetic field
\citep{Pattle_2017}. Another possible explanation is that OMC-1 is at a later evolutionary stage \citep{Castets_1995}. After the long-term evolution, the occurrence of gravitational collapse would be restricted to along field lines, and the perpendicular alignment implies that the magnetic field cannot efficiently resist gravity \citep{Koch_2018}. Compared with OMC-2/3, gravity plays a more important role in OMC-1. The results suggest that perpendicular alignment may only exist in well-evolved, gravity-dominated regions, but observational evidence is still lacking. Further research with larger samples is needed in the future.

\subsection{Comparison of HROs with previous studies}
We find a similar slope at different scales using HRO analysis. The transition column density, representing the shift from parallel to perpendicular alignment, varies from $10^{21}$ to $4\times 10^{21} \ \mathrm{cm^{-2}}$ on large scales and from $8 \times 10^{22}$ to $2\times 10^{23} \ \mathrm{cm^{-2}}$ on small scales, depending on the tracers employed. The role of magnetic field is complex around star-forming regions, with a turning point in $\xi$ and AM \citep{Planck_2016c, Ching_2022}. After removing the data point of highest column density, the derived transition column density is approximately $4 \times 10^{21} \ \mathrm{cm^{-2}}$ at large scales and $1 \times 10^{23} \ \mathrm{cm^{-2}}$ at small scales. 

\begin{table*}
        \centering
        \caption{Summary of HRO analysis}
        \label{HRO_analysis}
        \begin{tabular}{cccccc} 
                \hline
                \hline
Source &  Distance & Resolution & Scale & $X_{\mathrm{HRO}}$& Reference \\

 & (pc) & ($^{\prime\prime}$)  & (pc) &  &  \\
\hline
Taurus & 140 &  600   &  0.42    &  21.84  & \citealt{Planck_2016c}  \\
Ophiuchus & 140  & 600 &  0.42   &  22.70 & \citealt{Planck_2016c} \\
Lupus & 140  &600  &  0.42    &   21.72  & \citealt{Planck_2016c}\\
Chamaeleon-Musca & 160 & 600  & 0.48  &  21.67 & \citealt{Planck_2016c}\\
Corona Australia (CrA) & 170  & 600 &  0.51  &  24.14 & \citealt{Planck_2016c}\\
Aquila Rift & 260 & 600  & 0.78  &  22.23 & \citealt{Planck_2016c} \\
Perseus & 300  & 600  &  0.90  & 21.76 & \citealt{Planck_2016c} \\
IC5146 & 400 & 600  & 1.20  &  21.79 & \citealt{Planck_2016c} \\
Cepheus & 440 & 600  & 1.32  &  21.90 & \citealt{Planck_2016c}\\
Orion & 450 & 600   & 1.35  &  21.88  & \citealt{Planck_2016c} \\
Vela C-South-Nest & 933 & 180   & 0.84 & 22.23 & \citealt{Soler_2017}\\
Vela C-South-Ridge & 933 & 180   & 0.84 & 22.40 & \citealt{Soler_2017}\\
Vela C-Centre-Nest & 933 & 180   & 0.84 & 22.62 & \citealt{Soler_2017}\\
Vela C-Centre-Ridge & 933 & 180   & 0.84 & 20.69 & \citealt{Soler_2017}\\
DR21 & 1400 & 14 &  0.10 & 21.20 & \citealt{Ching_2022}\\
Serpens Main & 415 & 14 & 0.03 & ----- & \citealt{Kwon_2022} \\
OrionA & 414 & 300 & 0.62 & 20.97-21.62 & this work \\
OMC-1 & 414& 20 & 0.04 & 22.92-23.24 & this work \\
OMC-2/3 & 414 & 20 & 0.04 & ----- & this work \\

\hline
        \end{tabular}

\end{table*}

\citet{Planck_2016c} investigated the HROs in ten nearby clouds (d $<$ 450 pc) with a resolution of $10^{\prime}$. These authors find a common transition from parallel to perpendicular with increasing column density, and the typical transition column density is approximately $5 \times 10^{21} \ \mathrm{cm^{-2}}$. \citet{Soler_2017} adopted the HRO analysis in the Vela C region (d $\sim$ 933 pc) with a resolution of $3^{\prime}$ using the BLASTPol balloon-borne telescope. The similar trend is found and the transition column density ranges from $ 5 \times 10^{21}$ to $ 2 \times 10^{23} \ \mathrm{cm^{-2}}$. 

\begin{figure*}[ht]
    \centering
    \begin{minipage}{0.49\textwidth}
        \centering
        \includegraphics[width=\textwidth]{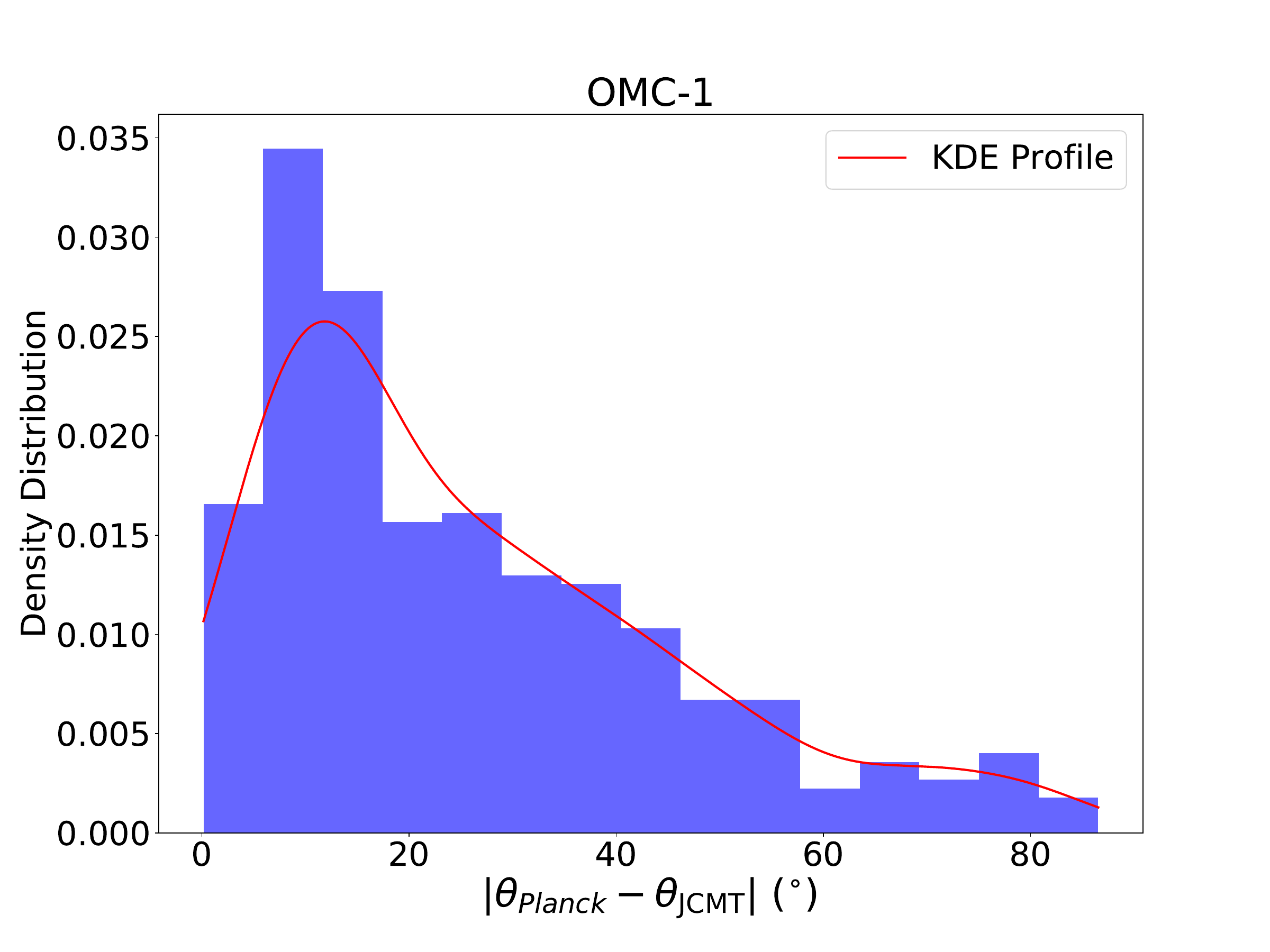} 
    \end{minipage}\hfill
    \begin{minipage}{0.49\textwidth}
        \centering
        \includegraphics[width=\textwidth]{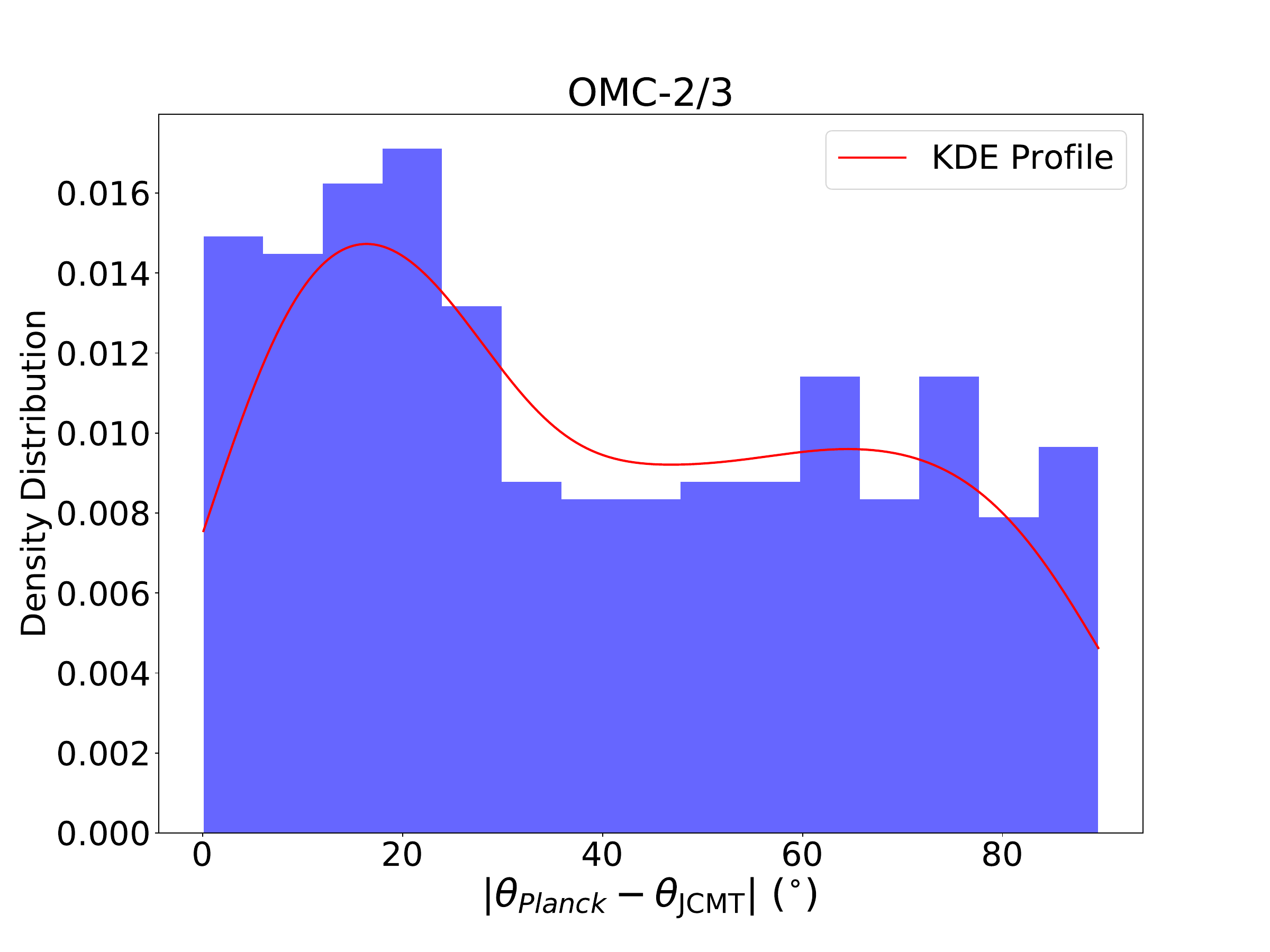} 
    \end{minipage}\hfill
\caption{Histogram of relative orientation between {\it{Planck-}} and JCMT-derived magnetic field in OMC-1 and OMC-2/3. The red line represents the Gaussian kernel density estimate of the angle difference.}
\label{kde}
\end{figure*}

On small scales, \citet{Ching_2022} applied the HRO analysis toward the DR21 filament (d $\sim$ 1.4 kpc) using JCMT with a resolution of $14^{\prime\prime}$ and the resulting transition column density is $1.6\times 10^{21} \ \mathrm{cm^{-2}}$. \citet{Kwon_2022} investigated the HRO of the Serpens Main region (d $\sim$ 415 pc) using JCMT with a resolution of $14^{\prime\prime}$. These latter authors find a complex relationship between $\xi$ and column density, which features several turning points and no clear transition column density. Detailed information regarding the entire HRO analysis is summarized in Table \ref{HRO_analysis}.

In summary, a consistent trend from parallel to perpendicular alignment with increasing column density is observed in most regions. However, the transition column density is scale-, tracer-, and environment-dependent. We do not find a uniform transition column density, suggesting that the HRO analysis should be applied case-by-case. We note that the comparison in Table \ref{HRO_analysis} is between single region studies (this work, \citealt{Soler_2017}, \citealt{Ching_2022}, \citealt{Kwon_2022}) and statistical studies (\citealt{Planck_2016c}). Conclusions from case studies cannot be directly applied to other star-forming regions.

The variation in relative orientation among different gas tracers is also interesting. \citet{Fissel_2019ApJ...878..110F} report no changes in relative orientation with increasing column density across all tracers. However, these authors observe a transition from parallel to perpendicular alignment when changing from low-density tracers to high-density tracers. Using a three-dimensional, turbulent collapsing-cloud MHD simulation, \citet{Mazzei} compared relative alignment between magnetic fields and molecular gas structure across different tracers. These authors find good agreement between their simulations and the results obtained in \citet{Fissel_2019ApJ...878..110F}. However, they suggest $^{12}$CO would remain in parallel alignment across the whole observer space, which is in contrast with our results. In our analysis, we not only observed a variation in relative orientation with different gas tracers but also identified changes in orientation with column density at different scales. This implies that the physical processes in the Orion A molecular cloud are hierarchical and highly complicated, making their explanation with simulations  challenging.

\subsection{Relative orientation between {{Planck}-} and JCMT- derived magnetic field}
Another interesting question to explore is the relative orientation between magnetic fields at different scales. \cite{Zhang_2014} first investigated this question by comparing SMA and parsec-scale magnetic fields and discovering a bimodal orientation distribution. These authors suggested that the magnetic field at the core scale could align either parallel or perpendicular to that of the clump scale. \cite{Li_2015} revealed that the magnetic field directions do not change significantly from cloud to clump and core scales in NGC 6334. In our study, we interpolate the {\it{Planck}} data and calculate the magnetic field orientation at large scales using the transformation in Sect. \ref{Planck_section}, and then compute the angular difference in specific regions. 

Figure \ref{kde} illustrates the distribution of angle differences between large-scale and small-scale magnetic fields. We use the Gaussian kernel density estimate (KDE) to represent the probability density function of the angle difference. In the OMC-1 region, small-scale magnetic fields align well with {\it{Planck}} data, with 79$\%$ of them showing an angle difference within 40 degrees of the large-scale magnetic fields. Regarding the OMC-2/3 region, the angle difference shows a broader distribution. This region exhibits a bimodal orientation distribution, which is similar to the findings reported by \cite{Zhang_2014}, with 59$\%$ of them displaying an angle difference within 40 degrees. The discrepancy in alignment between large-scale and small-scale magnetic fields could be attributed to the impact of foreground dust emission from Planck observations, as suggested by  \cite{Gu_2019}. A more important factor is the geometry of small-scale magnetic fields. As discussed in \cite{Zhang_2014}, a random orientation of fields would lead to a uniform distribution of angle differences. As the magnetic field properties vary significantly between the different subregions of OMC-2/3 \citep{Poidevin_2010}, the orientation distribution is much flatter than that in OMC-1. Whether or not star-forming activity affects the alignment of magnetic fields at different scales remains unclear, and further investigation with a larger sample and well-ordered magnetic field geometry is required in order to elucidate this topic.

\section{Summary}
\label{summary}

We compared the gas intensity structures of three gas tracers ($^{12}$CO, $^{13}$CO and C$^{18}$O) with magnetic fields at different scales. We applied the HRO technique in our analysis to examine the relation between the relative orientation of gas structures with respect to magnetic fields and column densities. Our main findings are as follows:

1. We find a similar trend in relative orientation between magnetic field and molecular gas intensity structures with respect to column densities at different scales in Orion A. Through a comparison with previous studies, we find the trend is common in molecular clouds at parsec scales but is not uniform at $<$ 0.1 pc scales. The relation between gas structures and magnetic field should be discussed case by case in individual star-forming regions at small scales.

2. When the orientation is not randomly distributed, we see a significant change in the relative orientation distribution between gas intensity structures and magnetic fields for three different molecular gas tracers. The perpendicular alignment is more clear when changing from low-density tracers to high-density tracers. The correlation between gas intensity structures and magnetic field is better traced by low-density gas tracers on large scales and high-density tracers on small scales, indicating the different roles of magnetic field in cloud formation and filament formation. The results found in this work only apply to the Orion A region and cannot be directly extrapolated to other star-forming regions.

\begin{acknowledgements}
We are grateful to an anonymous referee for the constructive comments that helped us improve this
paper.
      We would like to express our sincere gratitude to Qizhou Zhang, Tao-Chung Ching, Hua-bai Li, Juan Diego Soler, and Mengke Zhao for their valuable discussions and insightful suggestions that greatly improved this work.
This work has been supported by the National Science Foundation of China (12041305),
the China Manned Space Project (CMS-CSST-2021-A09, CMS-CSST-2021-B06),
and the China-Chile Joint Research Fund (CCJRF No. 2211). CCJRF is provided by Chinese Academy of Sciences South America Center for Astronomy (CASSACA) and established by National Astronomical Observatories, Chinese Academy of Sciences (NAOC) and Chilean Astronomy Society (SOCHIAS) to support China-Chile collaborations in astronomy. We acknowledge support from the Tianchi Talent Program of Xinjiang Uygur Autonomous Region.
      
\end{acknowledgements}


%

\vspace{5mm}





\bibliographystyle{aa}
\bibliography{sample63}

\begin{thebibliography}{70}
\expandafter\ifx\csname natexlab\endcsname\relax\def\natexlab#1{#1}\fi

\bibitem[{Andersson {et~al.}(2015)Andersson, Lazarian, \&
  Vaillancourt}]{Anderson_2015_review}
Andersson, B.-G., Lazarian, A., \& Vaillancourt, J.~E. 2015, Annual Review of
  Astronomy and Astrophysics, 53, 501

\bibitem[{{Andr{\'e}} {et~al.}(2022){Andr{\'e}}, {Palmeirim}, \&
  {Arzoumanian}}]{Andre_2022}
{Andr{\'e}}, P.~J., {Palmeirim}, P., \& {Arzoumanian}, D. 2022, \aap, 667, L1

\bibitem[{{Arzoumanian} {et~al.}(2019){Arzoumanian}, {Andr{\'e}},
  {K{\"o}nyves}, {Palmeirim}, {Roy}, {Schneider}, {Benedettini}, {Didelon}, {Di
  Francesco}, {Kirk}, \& {Ladjelate}}]{Arzoumanian_2019}
{Arzoumanian}, D., {Andr{\'e}}, P., {K{\"o}nyves}, V., {et~al.} 2019, \aap,
  621, A42

\bibitem[{{Arzoumanian} {et~al.}(2021){Arzoumanian}, {Furuya}, {Hasegawa},
  {Tahani}, {Sadavoy}, {Hull}, {Johnstone}, {Koch}, {Inutsuka}, {Doi}, {Hoang},
  {Onaka}, {Iwasaki}, {Shimajiri}, {Inoue}, {Peretto}, {Andr{\'e}}, {Bastien},
  {Berry}, {Chen}, {Di Francesco}, {Eswaraiah}, {Fanciullo}, {Fissel}, {Hwang},
  {Kang}, {Kim}, {Kim}, {Kirchschlager}, {Kwon}, {Lee}, {Liu}, {Lyo}, {Pattle},
  {Soam}, {Tang}, {Whitworth}, {Ching}, {Coud{\'e}}, {Wang}, {Ward-Thompson},
  {Lai}, {Qiu}, {Bourke}, {Byun}, {Chen}, {Chen}, {Chen}, {Cho}, {Choi},
  {Choi}, {Chrysostomou}, {Chung}, {Dai}, {Diep}, {Duan}, {Duan}, {Eden},
  {Fiege}, {Franzmann}, {Friberg}, {Fuller}, {Gledhill}, {Graves}, {Greaves},
  {Griffin}, {Gu}, {Han}, {Hatchell}, {Hayashi}, {Houde}, {Jeong}, {Kang},
  {Kang}, {Kataoka}, {Kawabata}, {Kemper}, {Kim}, {Kim}, {Kim}, {Kim}, {Kirk},
  {Kobayashi}, {K{\"o}nyves}, {Kusune}, {Kwon}, {Lacaille}, {Law}, {Lee},
  {Lee}, {Lee}, {Lee}, {Lee}, {Li}, {Li}, {Li}, {Liu}, {Liu}, {Liu}, {Lu},
  {Mairs}, {Matsumura}, {Matthews}, {Moriarty-Schieven}, {Nagata}, {Nakamura},
  {Nakanishi}, {Ngoc}, {Ohashi}, {Park}, {Parsons}, {Pyo}, {Qian}, {Rao},
  {Rawlings}, {Rawlings}, {Retter}, {Richer}, {Rigby}, {Saito}, {Savini},
  {Scaife}, {Seta}, {Shinnaga}, {Tamura}, {Tang}, {Tomisaka}, {Tram},
  {Tsukamoto}, {Viti}, {Wang}, {Xie}, {Yen}, {Yoo}, {Yuan}, {Yun}, {Zenko},
  {Zhang}, {Zhang}, {Zhang}, {Zhou}, {Zhu}, {de Looze}, {Dowell}, {Eyres},
  {Falle}, {Friesen}, {Robitaille}, \& {van Loo}}]{Arzoumanian_2021}
{Arzoumanian}, D., {Furuya}, R.~S., {Hasegawa}, T., {et~al.} 2021, \aap, 647,
  A78

\bibitem[{{Bally}(2008)}]{Bally_2008}
{Bally}, J. 2008, {Overview of the Orion Complex}, ed. B.~{Reipurth}, Vol.~4,
  459

\bibitem[{{Baug} {et~al.}(2021){Baug}, {Wang}, {Liu}, {Wu}, {Li}, {Zhang},
  {Tang}, {Goldsmith}, {Liu}, {Tej}, {Bronfman}, {Kim}, {Li}, {Lee},
  {Tatematsu}, {Hirota}, \& {Toth}}]{Baug_2021}
{Baug}, T., {Wang}, K., {Liu}, T., {et~al.} 2021, \mnras, 507, 4316

\bibitem[{{Bergin} \& {Tafalla}(2007)}]{2007ARA&A..45..339B}
{Bergin}, E.~A. \& {Tafalla}, M. 2007, \araa, 45, 339

\bibitem[{{Beuther} {et~al.}(2020){Beuther}, {Soler}, {Linz}, {Henning},
  {Gieser}, {Kuiper}, {Vlemmings}, {Hennebelle}, {Feng}, {Smith}, \&
  {Ahmadi}}]{Beuther_2020}
{Beuther}, H., {Soler}, J.~D., {Linz}, H., {et~al.} 2020, \apj, 904, 168

\bibitem[{{Castets} \& {Langer}(1995)}]{Castets_1995}
{Castets}, A. \& {Langer}, W.~D. 1995, \aap, 294, 835

\bibitem[{{Chandrasekhar} \& {Fermi}(1953)}]{CF}
{Chandrasekhar}, S. \& {Fermi}, E. 1953, \apj, 118, 113

\bibitem[{{Chen} {et~al.}(2020){Chen}, {Behrens}, {Washington}, {Fissel},
  {Friesen}, {Li}, {Pineda}, {Ginsburg}, {Kirk}, {Scibelli}, {Alves},
  {Redaelli}, {Caselli}, {Punanova}, {Di Francesco}, {Rosolowsky}, {Offner},
  {Martin}, {Chac{\'o}n-Tanarro}, {Chen}, {Chen}, {Keown}, {Seo}, {Shirley},
  {Arce}, {Goodman}, {Matzner}, {Myers}, \& {Singh}}]{Chen_2020}
{Chen}, C.-Y., {Behrens}, E.~A., {Washington}, J.~E., {et~al.} 2020, \mnras,
  494, 1971

\bibitem[{{Chen} {et~al.}(2016){Chen}, {King}, \& {Li}}]{Chen_2016}
{Chen}, C.-Y., {King}, P.~K., \& {Li}, Z.-Y. 2016, \apj, 829, 84

\bibitem[{{Chen} {et~al.}(2022){Chen}, {Li}, {Mazzei}, {Park}, {Fissel},
  {Chen}, {Klein}, \& {Li}}]{Chen_2022}
{Chen}, C.-Y., {Li}, Z.-Y., {Mazzei}, R.~R., {et~al.} 2022, \mnras, 514, 1575

\bibitem[{{Ching} {et~al.}(2022){Ching}, {Qiu}, {Li}, {Ren}, {Lai}, {Berry},
  {Pattle}, {Furuya}, {Ward-Thompson}, {Johnstone}, {Koch}, {Lee}, {Hoang},
  {Hasegawa}, {Kwon}, {Bastien}, {Eswaraiah}, {Wang}, {Kim}, {Hwang}, {Soam},
  {Lyo}, {Liu}, {Le Gouellec}, {Arzoumanian}, {Whitworth}, {Di Francesco},
  {Poidevin}, {Liu}, {Coud{\'e}}, {Tahani}, {Liu}, {Onaka}, {Li}, {Tamura},
  {Chen}, {Tang}, {Kirchschlager}, {Bourke}, {Byun}, {Chen}, {Chen}, {Chen},
  {Cho}, {Choi}, {Choi}, {Choi}, {Chrysostomou}, {Chung}, {Dai}, {Diep}, {Doi},
  {Duan}, {Duan}, {Eden}, {Fanciullo}, {Fiege}, {Fissel}, {Franzmann},
  {Friberg}, {Friesen}, {Fuller}, {Gledhill}, {Graves}, {Greaves}, {Griffin},
  {Gu}, {Han}, {Hayashi}, {Houde}, {Hull}, {Inoue}, {Inutsuka}, {Iwasaki},
  {Jeong}, {K{\"o}nyves}, {Kang}, {Kang}, {Karoly}, {Kataoka}, {Kawabata},
  {Kemper}, {Kim}, {Kim}, {Kim}, {Kim}, {Kim}, {Kim}, {Kirk}, {Kobayashi},
  {Kusune}, {Kwon}, {Lacaille}, {Law}, {Lee}, {Lee}, {Lee}, {Lee}, {Lee}, {Li},
  {Li}, {Lin}, {Liu}, {Lu}, {Mairs}, {Matsumura}, {Matthews},
  {Moriarty-Schieven}, {Nagata}, {Nakamura}, {Nakanishi}, {Ngoc}, {Ohashi},
  {Park}, {Parsons}, {Peretto}, {Priestley}, {Pyo}, {Qian}, {Rao}, {Rawlings},
  {Rawlings}, {Retter}, {Richer}, {Rigby}, {Sadavoy}, {Saito}, {Savini},
  {Seta}, {Shimajiri}, {Shinnaga}, {Tang}, {Tomisaka}, {Tram}, {Tsukamoto},
  {Viti}, {Wang}, {Wu}, {Xie}, {Yang}, {Yen}, {Yoo}, {Yuan}, {Yun}, {Zenko},
  {Zhang}, {Zhang}, {Zhang}, {Zhou}, {Zhu}, {de Looze}, {Andr{\'e}}, {Dowell},
  {Eyres}, {Falle}, {Robitaille}, \& {van Loo}}]{Ching_2022}
{Ching}, T.-C., {Qiu}, K., {Li}, D., {et~al.} 2022, \apj, 941, 122

\bibitem[{{Chini} {et~al.}(1997){Chini}, {Reipurth}, {Ward-Thompson}, {Bally},
  {Nyman}, {Sievers}, \& {Billawala}}]{Chini_1997}
{Chini}, R., {Reipurth}, B., {Ward-Thompson}, D., {et~al.} 1997, \apjl, 474,
  L135

\bibitem[{{Clark} \& {Johnson}(1974)}]{Clark_1974}
{Clark}, F.~O. \& {Johnson}, D.~R. 1974, \apjl, 191, L87

\bibitem[{{Corradi} {et~al.}(1998){Corradi}, {Aznar}, \&
  {Mampaso}}]{Corradi_1998}
{Corradi}, R. L.~M., {Aznar}, R., \& {Mampaso}, A. 1998, \mnras, 297, 617

\bibitem[{{Crutcher}(2012)}]{2012_Crutcher_ARA&A..50...29C}
{Crutcher}, R.~M. 2012, \araa, 50, 29

\bibitem[{{Crutcher} {et~al.}(1996){Crutcher}, {Troland}, {Lazareff}, \&
  {Kazes}}]{1996ApJ...456..217C}
{Crutcher}, R.~M., {Troland}, T.~H., {Lazareff}, B., \& {Kazes}, I. 1996, \apj,
  456, 217

\bibitem[{{Davis}(1951)}]{Davis}
{Davis}, L. 1951, Physical Review, 81, 890

\bibitem[{{Falgarone} {et~al.}(2008){Falgarone}, {Troland}, {Crutcher}, \&
  {Paubert}}]{2008_CN_Zeeman_A&A...487..247F}
{Falgarone}, E., {Troland}, T.~H., {Crutcher}, R.~M., \& {Paubert}, G. 2008,
  \aap, 487, 247

\bibitem[{{Fissel} {et~al.}(2019){Fissel}, {Ade}, {Angil{\`e}}, {Ashton},
  {Benton}, {Chen}, {Cunningham}, {Devlin}, {Dober}, {Friesen}, {Fukui},
  {Galitzki}, {Gandilo}, {Goodman}, {Green}, {Jones}, {Klein}, {King},
  {Korotkov}, {Li}, {Lowe}, {Martin}, {Matthews}, {Moncelsi}, {Nakamura},
  {Netterfield}, {Newmark}, {Novak}, {Pascale}, {Poidevin}, {Santos}, {Savini},
  {Scott}, {Shariff}, {Soler}, {Thomas}, {Tucker}, {Tucker}, {Ward-Thompson},
  \& {Zucker}}]{Fissel_2019ApJ...878..110F}
{Fissel}, L.~M., {Ade}, P. A.~R., {Angil{\`e}}, F.~E., {et~al.} 2019, \apj,
  878, 110

\bibitem[{{Frerking} {et~al.}(1982){Frerking}, {Langer}, \&
  {Wilson}}]{Frerking_1982ApJ...262..590F}
{Frerking}, M.~A., {Langer}, W.~D., \& {Wilson}, R.~W. 1982, \apj, 262, 590

\bibitem[{Gonz{\'{a}}lez-Casanova \& Lazarian(2017)}]{Gonz_lez_Casanova_2017}
Gonz{\'{a}}lez-Casanova, D.~F. \& Lazarian, A. 2017, The Astrophysical Journal,
  835, 41

\bibitem[{{Gu} \& {Li}(2019)}]{Gu_2019}
{Gu}, Q. \& {Li}, H.-b. 2019, \apjl, 871, L15

\bibitem[{{Hennebelle} \& {Falgarone}(2012)}]{2012A&ARv..20...55H}
{Hennebelle}, P. \& {Falgarone}, E. 2012, \aapr, 20, 55

\bibitem[{{Houde} {et~al.}(2004){Houde}, {Dowell}, {Hildebrand}, {Dotson},
  {Vaillancourt}, {Phillips}, {Peng}, \& {Bastien}}]{Houde_2004}
{Houde}, M., {Dowell}, C.~D., {Hildebrand}, R.~H., {et~al.} 2004, \apj, 604,
  717

\bibitem[{Jow {et~al.}(2017)Jow, Hill, Scott, Soler, Martin, Devlin, Fissel, \&
  Poidevin}]{Jow_2018}
Jow, D.~L., Hill, R., Scott, D., {et~al.} 2017, Monthly Notices of the Royal
  Astronomical Society, 474, 1018

\bibitem[{{Koch} {et~al.}(2013){Koch}, {Tang}, \& {Ho}}]{Koch_2013}
{Koch}, P.~M., {Tang}, Y.-W., \& {Ho}, P. T.~P. 2013, \apj, 775, 77

\bibitem[{{Koch} {et~al.}(2018){Koch}, {Tang}, {Ho}, {Yen}, {Su}, \&
  {Takakuwa}}]{Koch_2018}
{Koch}, P.~M., {Tang}, Y.-W., {Ho}, P. T.~P., {et~al.} 2018, \apj, 855, 39

\bibitem[{{Kwon} {et~al.}(2022){Kwon}, {Pattle}, {Sadavoy}, {Hull},
  {Johnstone}, {Ward-Thompson}, {Di Francesco}, {Koch}, {Furuya}, {Doi}, {Le
  Gouellec}, {Hwang}, {Lyo}, {Soam}, {Tang}, {Hoang}, {Kirchschlager},
  {Eswaraiah}, {Fanciullo}, {Kim}, {Onaka}, {K{\"o}nyves}, {Kang}, {Lee},
  {Tamura}, {Bastien}, {Hasegawa}, {Lai}, {Qiu}, {Berry}, {Arzoumanian},
  {Bourke}, {Byun}, {Chen}, {Chen}, {Chen}, {Chen}, {Ching}, {Cho}, {Choi},
  {Choi}, {Chrysostomou}, {Chung}, {Coud{\'e}}, {Dai}, {Diep}, {Duan}, {Duan},
  {Eden}, {Fiege}, {Fissel}, {Franzmann}, {Friberg}, {Friesen}, {Fuller},
  {Gledhill}, {Graves}, {Greaves}, {Griffin}, {Gu}, {Han}, {Hatchell},
  {Hayashi}, {Houde}, {Inoue}, {Inutsuka}, {Iwasaki}, {Jeong}, {Kang},
  {Karoly}, {Kataoka}, {Kawabata}, {Kemper}, {Kim}, {Kim}, {Kim}, {Kim}, {Kim},
  {Kirk}, {Kobayashi}, {Kusune}, {Kwon}, {Lacaille}, {Law}, {Lee}, {Lee},
  {Lee}, {Lee}, {Lee}, {Li}, {Li}, {Li}, {Lin}, {Liu}, {Liu}, {Liu}, {Liu},
  {Lu}, {Mairs}, {Matsumura}, {Matthews}, {Moriarty-Schieven}, {Nagata},
  {Nakamura}, {Nakanishi}, {Ngoc}, {Ohashi}, {Park}, {Parsons}, {Peretto},
  {Priestley}, {Pyo}, {Qian}, {Rao}, {Rawlings}, {Rawlings}, {Retter},
  {Richer}, {Rigby}, {Saito}, {Savini}, {Seta}, {Shimajiri}, {Shinnaga},
  {Tahani}, {Tang}, {Tomisaka}, {Tram}, {Tsukamoto}, {Viti}, {Wang}, {Wang},
  {Whitworth}, {Wu}, {Xie}, {Yen}, {Yoo}, {Yuan}, {Yun}, {Zenko}, {Zhang},
  {Zhang}, {Zhang}, {Zhou}, {Zhu}, {de Looze}, {Andr{\'e}}, {Dowell}, {Eyres},
  {Falle}, {Robitaille}, \& {Loo}}]{Kwon_2022}
{Kwon}, W., {Pattle}, K., {Sadavoy}, S., {et~al.} 2022, \apj, 926, 163

\bibitem[{{Lamarre} {et~al.}(2010){Lamarre}, {Puget}, {Ade}, {Bouchet},
  {Guyot}, {Lange}, {Pajot}, {Arondel}, {Benabed}, {Beney}, {Beno{\^\i}t},
  {Bernard}, {Bhatia}, {Blanc}, {Bock}, {Br{\'e}elle}, {Bradshaw}, {Camus},
  {Catalano}, {Charra}, {Charra}, {Church}, {Couchot}, {Coulais}, {Crill},
  {Crook}, {Dassas}, {de Bernardis}, {Delabrouille}, {de Marcillac}, {Delouis},
  {D{\'e}sert}, {Dumesnil}, {Dupac}, {Efstathiou}, {Eng}, {Evesque},
  {Fourmond}, {Ganga}, {Giard}, {Gispert}, {Guglielmi}, {Haissinski},
  {Henrot-Versill{\'e}}, {Hivon}, {Holmes}, {Jones}, {Koch}, {Lagard{\`e}re},
  {Lami}, {Land{\'e}}, {Leriche}, {Leroy}, {Longval}, {Mac{\'\i}as-P{\'e}rez},
  {Maciaszek}, {Maffei}, {Mansoux}, {Marty}, {Masi}, {Mercier},
  {Miville-Desch{\^e}nes}, {Moneti}, {Montier}, {Murphy}, {Narbonne}, {Nexon},
  {Paine}, {Pahn}, {Perdereau}, {Piacentini}, {Piat}, {Plaszczynski},
  {Pointecouteau}, {Pons}, {Ponthieu}, {Prunet}, {Rambaud}, {Recouvreur},
  {Renault}, {Ristorcelli}, {Rosset}, {Santos}, {Savini}, {Serra}, {Stassi},
  {Sudiwala}, {Sygnet}, {Tauber}, {Torre}, {Tristram}, {Vibert}, {Woodcraft},
  {Yurchenko}, \& {Yvon}}]{Lamarre_2010}
{Lamarre}, J.~M., {Puget}, J.~L., {Ade}, P.~A.~R., {et~al.} 2010, \aap, 520, A9

\bibitem[{{Lazarian} \& {Yuen}(2018)}]{Lazarian_2018}
{Lazarian}, A. \& {Yuen}, K.~H. 2018, \apj, 853, 96

\bibitem[{{Li} {et~al.}(2018){Li}, {Wang}, {Zhang}, {Ma}, {Fang}, \&
  {Yang}}]{Li_2018}
{Li}, C., {Wang}, H., {Zhang}, M., {et~al.} 2018, \apjs, 238, 10

\bibitem[{{Li} {et~al.}(2015){Li}, {Yuen}, {Otto}, {Leung}, {Sridharan},
  {Zhang}, {Liu}, {Tang}, \& {Qiu}}]{Li_2015}
{Li}, H.-B., {Yuen}, K.~H., {Otto}, F., {et~al.} 2015, \nat, 520, 518

\bibitem[{{Liu} {et~al.}(2021){Liu}, {Zhang}, {Commer{\c{c}}on}, {Valdivia},
  {Maury}, \& {Qiu}}]{Liu_2021}
{Liu}, J., {Zhang}, Q., {Commer{\c{c}}on}, B., {et~al.} 2021, \apj, 919, 79

\bibitem[{{Liu} {et~al.}(2023){Liu}, {Zhang}, {Koch}, {Liu}, {Li}, {Li},
  {Girart}, {Chen}, {Ching}, {Ho}, {Lai}, {Qiu}, {Rao}, \& {Tang}}]{Liu_2023}
{Liu}, J., {Zhang}, Q., {Koch}, P.~M., {et~al.} 2023, \apj, 945, 160

\bibitem[{{Liu} {et~al.}(2022){Liu}, {Zhang}, \& {Qiu}}]{Liu_2022}
{Liu}, J., {Zhang}, Q., \& {Qiu}, K. 2022, Frontiers in Astronomy and Space
  Sciences, 9, 943556

\bibitem[{{Mangum} \& {Shirley}(2015)}]{Mangum&shirley_2015}
{Mangum}, J.~G. \& {Shirley}, Y.~L. 2015, \pasp, 127, 266

\bibitem[{Matthews {et~al.}(2009)Matthews, McPhee, Fissel, \&
  Curran}]{Matthews_2009}
Matthews, B.~C., McPhee, C.~A., Fissel, L.~M., \& Curran, R.~L. 2009, The
  Astrophysical Journal Supplement Series, 182, 143

\bibitem[{{Mazzei} {et~al.}(2023){Mazzei}, {Li}, {Chen}, {Fissel}, {Chen}, \&
  {Park}}]{Mazzei}
{Mazzei}, R., {Li}, Z.-Y., {Chen}, C.-Y., {et~al.} 2023, \mnras, 521, 3830

\bibitem[{{McKee} \& {Ostriker}(2007)}]{2007ARA&A..45..565M}
{McKee}, C.~F. \& {Ostriker}, E.~C. 2007, \araa, 45, 565

\bibitem[{{Menten} {et~al.}(2007){Menten}, {Reid}, {Forbrich}, \&
  {Brunthaler}}]{Menten_2007}
{Menten}, K.~M., {Reid}, M.~J., {Forbrich}, J., \& {Brunthaler}, A. 2007, \aap,
  474, 515

\bibitem[{{Mouschovias}(1976{\natexlab{a}})}]{Mouschovias_1976a}
{Mouschovias}, T.~C. 1976{\natexlab{a}}, \apj, 206, 753

\bibitem[{{Mouschovias}(1976{\natexlab{b}})}]{Mouschovias_1976b}
{Mouschovias}, T.~C. 1976{\natexlab{b}}, \apj, 207, 141

\bibitem[{{Nakamura} {et~al.}(2019){Nakamura}, {Ishii}, {Dobashi},
  {Shimoikura}, {Shimajiri}, {Kawabe}, {Tanabe}, {Hirose}, {Oyamada},
  {Urasawa}, {Takemura}, {Tsukagoshi}, {Momose}, {Sugitani}, {Nishi},
  {Okumura}, {Sanhueza}, {Nguyen-Luong}, \&
  {Kusune}}]{Nobeyama_star_forming_project_overview}
{Nakamura}, F., {Ishii}, S., {Dobashi}, K., {et~al.} 2019, \pasj, 71, S3

\bibitem[{{Pattle} {et~al.}(2017){Pattle}, {Ward-Thompson}, {Berry},
  {Hatchell}, {Chen}, {Pon}, {Koch}, {Kwon}, {Kim}, {Bastien}, {Cho},
  {Coud{\'e}}, {Di Francesco}, {Fuller}, {Furuya}, {Graves}, {Johnstone},
  {Kirk}, {Kwon}, {Lee}, {Matthews}, {Mottram}, {Parsons}, {Sadavoy},
  {Shinnaga}, {Soam}, {Hasegawa}, {Lai}, {Qiu}, \& {Friberg}}]{Pattle_2017}
{Pattle}, K., {Ward-Thompson}, D., {Berry}, D., {et~al.} 2017, \apj, 846, 122

\bibitem[{{Pillai} {et~al.}(2020){Pillai}, {Clemens}, {Reissl}, {Myers},
  {Kauffmann}, {Lopez-Rodriguez}, {Alves}, {Franco}, {Henshaw}, {Menten},
  {Nakamura}, {Seifried}, {Sugitani}, \& {Wiesemeyer}}]{Pillai_2020}
{Pillai}, T. G.~S., {Clemens}, D.~P., {Reissl}, S., {et~al.} 2020, Nature
  Astronomy, 4, 1195

\bibitem[{{Pineda} {et~al.}(2010){Pineda}, {Goldsmith}, {Chapman}, {Snell},
  {Li}, {Cambr{\'e}sy}, \& {Brunt}}]{Pineda_2010}
{Pineda}, J.~L., {Goldsmith}, P.~F., {Chapman}, N., {et~al.} 2010, \apj, 721,
  686

\bibitem[{{Planck Collaboration} {et~al.}(2016{\natexlab{a}}){Planck
  Collaboration}, {Adam}, {Ade}, {Aghanim}, {Alves}, {Arnaud}, {Arzoumanian},
  {Ashdown}, {Aumont}, {Baccigalupi}, {Banday}, {Barreiro}, {Bartolo},
  {Battaner}, {Benabed}, {Benoit-L{\'e}vy}, {Bernard}, {Bersanelli},
  {Bielewicz}, {Bonaldi}, {Bonavera}, {Bond}, {Borrill}, {Bouchet},
  {Boulanger}, {Bracco}, {Burigana}, {Butler}, {Calabrese}, {Cardoso},
  {Catalano}, {Chamballu}, {Chiang}, {Christensen}, {Colombi}, {Colombo},
  {Combet}, {Couchot}, {Crill}, {Curto}, {Cuttaia}, {Danese}, {Davies},
  {Davis}, {de Bernardis}, {de Rosa}, {de Zotti}, {Delabrouille}, {Dickinson},
  {Diego}, {Dole}, {Donzelli}, {Dor{\'e}}, {Douspis}, {Ducout}, {Dupac},
  {Efstathiou}, {Elsner}, {En{\ss}lin}, {Eriksen}, {Falgarone}, {Ferri{\`e}re},
  {Finelli}, {Forni}, {Frailis}, {Fraisse}, {Franceschi}, {Frejsel},
  {Galeotta}, {Galli}, {Ganga}, {Ghosh}, {Giard}, {Gjerl{\o}w},
  {Gonz{\'a}lez-Nuevo}, {G{\'o}rski}, {Gregorio}, {Gruppuso}, {Guillet},
  {Hansen}, {Hanson}, {Harrison}, {Henrot-Versill{\'e}},
  {Hern{\'a}ndez-Monteagudo}, {Herranz}, {Hildebrandt}, {Hivon}, {Hobson},
  {Holmes}, {Hovest}, {Huffenberger}, {Hurier}, {Jaffe}, {Jaffe}, {Jones},
  {Juvela}, {Keih{\"a}nen}, {Keskitalo}, {Kisner}, {Kneissl}, {Knoche}, {Kunz},
  {Kurki-Suonio}, {Lagache}, {Lamarre}, {Lasenby}, {Lattanzi}, {Lawrence},
  {Leonardi}, {Levrier}, {Liguori}, {Lilje}, {Linden-V{\o}rnle},
  {L{\'o}pez-Caniego}, {Lubin}, {Mac{\'\i}as-P{\'e}rez}, {Maffei}, {Maino},
  {Mandolesi}, {Maris}, {Marshall}, {Martin}, {Mart{\'\i}nez-Gonz{\'a}lez},
  {Masi}, {Matarrese}, {Mazzotta}, {Melchiorri}, {Mendes}, {Mennella},
  {Migliaccio}, {Miville-Desch{\^e}nes}, {Moneti}, {Montier}, {Morgante},
  {Mortlock}, {Munshi}, {Murphy}, {Naselsky}, {Natoli}, {N{\o}rgaard-Nielsen},
  {Noviello}, {Novikov}, {Novikov}, {Oppermann}, {Oxborrow}, {Pagano}, {Pajot},
  {Paoletti}, {Pasian}, {Perdereau}, {Perotto}, {Perrotta}, {Pettorino},
  {Piacentini}, {Piat}, {Plaszczynski}, {Pointecouteau}, {Polenta}, {Ponthieu},
  {Popa}, {Pratt}, {Prunet}, {Puget}, {Rachen}, {Reach}, {Reinecke},
  {Remazeilles}, {Renault}, {Ristorcelli}, {Rocha}, {Roudier},
  {Rubi{\~n}o-Mart{\'\i}n}, {Rusholme}, {Sandri}, {Santos}, {Savini}, {Scott},
  {Soler}, {Spencer}, {Stolyarov}, {Sudiwala}, {Sunyaev}, {Sutton},
  {Suur-Uski}, {Sygnet}, {Tauber}, {Terenzi}, {Toffolatti}, {Tomasi},
  {Tristram}, {Tucci}, {Umana}, {Valenziano}, {Valiviita}, {Van Tent},
  {Vielva}, {Villa}, {Wade}, {Wandelt}, {Wehus}, {Wiesemeyer}, {Yvon},
  {Zacchei}, \& {Zonca}}]{planck_HRO_diffuse}
{Planck Collaboration}, {Adam}, R., {Ade}, P.~A.~R., {et~al.}
  2016{\natexlab{a}}, \aap, 586, A135

\bibitem[{{Planck Collaboration} {et~al.}(2016{\natexlab{b}}){Planck
  Collaboration}, {Ade}, {Aghanim}, {Alves}, {Arnaud}, {Arzoumanian},
  {Ashdown}, {Aumont}, {Baccigalupi}, {Banday}, {Barreiro}, {Bartolo},
  {Battaner}, {Benabed}, {Beno{\^\i}t}, {Benoit-L{\'e}vy}, {Bernard},
  {Bersanelli}, {Bielewicz}, {Bock}, {Bonavera}, {Bond}, {Borrill}, {Bouchet},
  {Boulanger}, {Bracco}, {Burigana}, {Calabrese}, {Cardoso}, {Catalano},
  {Chiang}, {Christensen}, {Colombo}, {Combet}, {Couchot}, {Crill}, {Curto},
  {Cuttaia}, {Danese}, {Davies}, {Davis}, {de Bernardis}, {de Rosa}, {de
  Zotti}, {Delabrouille}, {Dickinson}, {Diego}, {Dole}, {Donzelli}, {Dor{\'e}},
  {Douspis}, {Ducout}, {Dupac}, {Efstathiou}, {Elsner}, {En{\ss}lin},
  {Eriksen}, {Falceta-Gon{\c{c}}alves}, {Falgarone}, {Ferri{\`e}re}, {Finelli},
  {Forni}, {Frailis}, {Fraisse}, {Franceschi}, {Frejsel}, {Galeotta}, {Galli},
  {Ganga}, {Ghosh}, {Giard}, {Gjerl{\o}w}, {Gonz{\'a}lez-Nuevo}, {G{\'o}rski},
  {Gregorio}, {Gruppuso}, {Gudmundsson}, {Guillet}, {Harrison}, {Helou},
  {Hennebelle}, {Henrot-Versill{\'e}}, {Hern{\'a}ndez-Monteagudo}, {Herranz},
  {Hildebrandt}, {Hivon}, {Holmes}, {Hornstrup}, {Huffenberger}, {Hurier},
  {Jaffe}, {Jaffe}, {Jones}, {Juvela}, {Keih{\"a}nen}, {Keskitalo}, {Kisner},
  {Knoche}, {Kunz}, {Kurki-Suonio}, {Lagache}, {Lamarre}, {Lasenby},
  {Lattanzi}, {Lawrence}, {Leonardi}, {Levrier}, {Liguori}, {Lilje},
  {Linden-V{\o}rnle}, {L{\'o}pez-Caniego}, {Lubin}, {Mac{\'\i}as-P{\'e}rez},
  {Maino}, {Mandolesi}, {Mangilli}, {Maris}, {Martin},
  {Mart{\'\i}nez-Gonz{\'a}lez}, {Masi}, {Matarrese}, {Melchiorri}, {Mendes},
  {Mennella}, {Migliaccio}, {Miville-Desch{\^e}nes}, {Moneti}, {Montier},
  {Morgante}, {Mortlock}, {Munshi}, {Murphy}, {Naselsky}, {Nati},
  {Netterfield}, {Noviello}, {Novikov}, {Novikov}, {Oppermann}, {Oxborrow},
  {Pagano}, {Pajot}, {Paladini}, {Paoletti}, {Pasian}, {Perotto}, {Pettorino},
  {Piacentini}, {Piat}, {Pierpaoli}, {Pietrobon}, {Plaszczynski},
  {Pointecouteau}, {Polenta}, {Ponthieu}, {Pratt}, {Prunet}, {Puget}, {Rachen},
  {Reinecke}, {Remazeilles}, {Renault}, {Renzi}, {Ristorcelli}, {Rocha},
  {Rossetti}, {Roudier}, {Rubi{\~n}o-Mart{\'\i}n}, {Rusholme}, {Sandri},
  {Santos}, {Savelainen}, {Savini}, {Scott}, {Soler}, {Stolyarov}, {Sudiwala},
  {Sutton}, {Suur-Uski}, {Sygnet}, {Tauber}, {Terenzi}, {Toffolatti}, {Tomasi},
  {Tristram}, {Tucci}, {Umana}, {Valenziano}, {Valiviita}, {Van Tent},
  {Vielva}, {Villa}, {Wade}, {Wandelt}, {Wehus}, {Ysard}, {Yvon}, \&
  {Zonca}}]{Planck_2016c}
{Planck Collaboration}, {Ade}, P.~A.~R., {Aghanim}, N., {et~al.}
  2016{\natexlab{b}}, \aap, 586, A138

\bibitem[{{Planck Collaboration} {et~al.}(2020{\natexlab{a}}){Planck
  Collaboration}, {Aghanim}, {Akrami}, {Arroja}, {Ashdown}, {Aumont},
  {Baccigalupi}, {Ballardini}, {Banday}, {Barreiro}, {Bartolo}, {Basak},
  {Battye}, {Benabed}, {Bernard}, {Bersanelli}, {Bielewicz}, {Bock}, {Bond},
  {Borrill}, {Bouchet}, {Boulanger}, {Bucher}, {Burigana}, {Butler},
  {Calabrese}, {Cardoso}, {Carron}, {Casaponsa}, {Challinor}, {Chiang},
  {Colombo}, {Combet}, {Contreras}, {Crill}, {Cuttaia}, {de Bernardis}, {de
  Zotti}, {Delabrouille}, {Delouis}, {D{\'e}sert}, {Di Valentino}, {Dickinson},
  {Diego}, {Donzelli}, {Dor{\'e}}, {Douspis}, {Ducout}, {Dupac}, {Efstathiou},
  {Elsner}, {En{\ss}lin}, {Eriksen}, {Falgarone}, {Fantaye}, {Fergusson},
  {Fernandez-Cobos}, {Finelli}, {Forastieri}, {Frailis}, {Franceschi},
  {Frolov}, {Galeotta}, {Galli}, {Ganga}, {G{\'e}nova-Santos}, {Gerbino},
  {Ghosh}, {Gonz{\'a}lez-Nuevo}, {G{\'o}rski}, {Gratton}, {Gruppuso},
  {Gudmundsson}, {Hamann}, {Handley}, {Hansen}, {Helou}, {Herranz},
  {Hildebrandt}, {Hivon}, {Huang}, {Jaffe}, {Jones}, {Karakci}, {Keih{\"a}nen},
  {Keskitalo}, {Kiiveri}, {Kim}, {Kisner}, {Knox}, {Krachmalnicoff}, {Kunz},
  {Kurki-Suonio}, {Lagache}, {Lamarre}, {Langer}, {Lasenby}, {Lattanzi},
  {Lawrence}, {Le Jeune}, {Leahy}, {Lesgourgues}, {Levrier}, {Lewis},
  {Liguori}, {Lilje}, {Lilley}, {Lindholm}, {L{\'o}pez-Caniego}, {Lubin}, {Ma},
  {Mac{\'\i}as-P{\'e}rez}, {Maggio}, {Maino}, {Mandolesi}, {Mangilli},
  {Marcos-Caballero}, {Maris}, {Martin}, {Martinelli},
  {Mart{\'\i}nez-Gonz{\'a}lez}, {Matarrese}, {Mauri}, {McEwen}, {Meerburg},
  {Meinhold}, {Melchiorri}, {Mennella}, {Migliaccio}, {Millea}, {Mitra},
  {Miville-Desch{\^e}nes}, {Molinari}, {Moneti}, {Montier}, {Morgante}, {Moss},
  {Mottet}, {M{\"u}nchmeyer}, {Natoli}, {N{\o}rgaard-Nielsen}, {Oxborrow},
  {Pagano}, {Paoletti}, {Partridge}, {Patanchon}, {Pearson}, {Peel}, {Peiris},
  {Perrotta}, {Pettorino}, {Piacentini}, {Polastri}, {Polenta}, {Puget},
  {Rachen}, {Reinecke}, {Remazeilles}, {Renault}, {Renzi}, {Rocha}, {Rosset},
  {Roudier}, {Rubi{\~n}o-Mart{\'\i}n}, {Ruiz-Granados}, {Salvati}, {Sandri},
  {Savelainen}, {Scott}, {Shellard}, {Shiraishi}, {Sirignano}, {Sirri},
  {Spencer}, {Sunyaev}, {Suur-Uski}, {Tauber}, {Tavagnacco}, {Tenti},
  {Terenzi}, {Toffolatti}, {Tomasi}, {Trombetti}, {Valiviita}, {Van Tent},
  {Vibert}, {Vielva}, {Villa}, {Vittorio}, {Wandelt}, {Wehus}, {White},
  {White}, {Zacchei}, \& {Zonca}}]{planck_2020a}
{Planck Collaboration}, {Aghanim}, N., {Akrami}, Y., {et~al.}
  2020{\natexlab{a}}, \aap, 641, A1

\bibitem[{{Planck Collaboration} {et~al.}(2020{\natexlab{b}}){Planck
  Collaboration}, {Aghanim}, {Akrami}, {Ashdown}, {Aumont}, {Baccigalupi},
  {Ballardini}, {Banday}, {Barreiro}, {Bartolo}, {Basak}, {Benabed}, {Bernard},
  {Bersanelli}, {Bielewicz}, {Bond}, {Borrill}, {Bouchet}, {Boulanger},
  {Bucher}, {Burigana}, {Calabrese}, {Cardoso}, {Carron}, {Challinor},
  {Chiang}, {Colombo}, {Combet}, {Couchot}, {Crill}, {Cuttaia}, {de Bernardis},
  {de Rosa}, {de Zotti}, {Delabrouille}, {Delouis}, {Di Valentino}, {Diego},
  {Dor{\'e}}, {Douspis}, {Ducout}, {Dupac}, {Efstathiou}, {Elsner},
  {En{\ss}lin}, {Eriksen}, {Falgarone}, {Fantaye}, {Finelli}, {Frailis},
  {Fraisse}, {Franceschi}, {Frolov}, {Galeotta}, {Galli}, {Ganga},
  {G{\'e}nova-Santos}, {Gerbino}, {Ghosh}, {Gonz{\'a}lez-Nuevo}, {G{\'o}rski},
  {Gratton}, {Gruppuso}, {Gudmundsson}, {Handley}, {Hansen},
  {Henrot-Versill{\'e}}, {Herranz}, {Hivon}, {Huang}, {Jaffe}, {Jones},
  {Karakci}, {Keih{\"a}nen}, {Keskitalo}, {Kiiveri}, {Kim}, {Kisner},
  {Krachmalnicoff}, {Kunz}, {Kurki-Suonio}, {Lagache}, {Lamarre}, {Lasenby},
  {Lattanzi}, {Lawrence}, {Levrier}, {Liguori}, {Lilje}, {Lindholm},
  {L{\'o}pez-Caniego}, {Ma}, {Mac{\'\i}as-P{\'e}rez}, {Maggio}, {Maino},
  {Mandolesi}, {Mangilli}, {Martin}, {Mart{\'\i}nez-Gonz{\'a}lez}, {Matarrese},
  {Mauri}, {McEwen}, {Melchiorri}, {Mennella}, {Migliaccio},
  {Miville-Desch{\^e}nes}, {Molinari}, {Moneti}, {Montier}, {Morgante}, {Moss},
  {Mottet}, {Natoli}, {Pagano}, {Paoletti}, {Partridge}, {Patanchon},
  {Patrizii}, {Perdereau}, {Perrotta}, {Pettorino}, {Piacentini}, {Puget},
  {Rachen}, {Reinecke}, {Remazeilles}, {Renzi}, {Rocha}, {Roudier}, {Salvati},
  {Sandri}, {Savelainen}, {Scott}, {Sirignano}, {Sirri}, {Spencer}, {Sunyaev},
  {Suur-Uski}, {Tauber}, {Tavagnacco}, {Tenti}, {Toffolatti}, {Tomasi},
  {Tristram}, {Trombetti}, {Valiviita}, {Vansyngel}, {Van Tent}, {Vibert},
  {Vielva}, {Villa}, {Vittorio}, {Wandelt}, {Wehus}, \& {Zonca}}]{Planck_2020c}
{Planck Collaboration}, {Aghanim}, N., {Akrami}, Y., {et~al.}
  2020{\natexlab{b}}, \aap, 641, A3

\bibitem[{{Plaszczynski} {et~al.}(2014){Plaszczynski}, {Montier}, {Levrier}, \&
  {Tristram}}]{Plaszczynski_2014}
{Plaszczynski}, S., {Montier}, L., {Levrier}, F., \& {Tristram}, M. 2014,
  \mnras, 439, 4048

\bibitem[{{Poidevin} {et~al.}(2010){Poidevin}, {Bastien}, \&
  {Matthews}}]{Poidevin_2010}
{Poidevin}, F., {Bastien}, P., \& {Matthews}, B.~C. 2010, \apj, 716, 893

\bibitem[{{Sanhueza} {et~al.}(2021){Sanhueza}, {Girart}, {Padovani}, {Galli},
  {Hull}, {Zhang}, {Cortes}, {Stephens}, {Fern{\'a}ndez-L{\'o}pez}, {Jackson},
  {Frau}, {Kock}, {Wu}, {Zapata}, {Olguin}, {Lu}, {Silva}, {Tang}, {Sakai},
  {Guzm{\'a}n}, {Tatematsu}, {Nakamura}, \& {Chen}}]{Sanhueza_2021}
{Sanhueza}, P., {Girart}, J.~M., {Padovani}, M., {et~al.} 2021, \apjl, 915, L10

\bibitem[{{Seifried} {et~al.}(2020){Seifried}, {Walch}, {Weis}, {Reissl},
  {Soler}, {Klessen}, \& {Joshi}}]{Seifried_2020}
{Seifried}, D., {Walch}, S., {Weis}, M., {et~al.} 2020, \mnras, 497, 4196

\bibitem[{{Skalidis} {et~al.}(2021){Skalidis}, {Sternberg}, {Beattie},
  {Pavlidou}, \& {Tassis}}]{Skalidis_2021b}
{Skalidis}, R., {Sternberg}, J., {Beattie}, J.~R., {Pavlidou}, V., \& {Tassis},
  K. 2021, \aap, 656, A118

\bibitem[{{Skalidis} \& {Tassis}(2021)}]{Skalidis_2021a}
{Skalidis}, R. \& {Tassis}, K. 2021, \aap, 647, A186

\bibitem[{Sokolov {et~al.}(2019)Sokolov, Wang, Pineda, Caselli, Henshaw,
  Barnes, Tan, Fontani, \& Jim{\'{e}}nez-Serra}]{Sokolov_2019}
Sokolov, V., Wang, K., Pineda, J.~E., {et~al.} 2019, The Astrophysical Journal,
  872, 30

\bibitem[{{Soler}(2019)}]{Soler_2019}
{Soler}, J.~D. 2019, \aap, 629, A96

\bibitem[{{Soler} {et~al.}(2017){Soler}, {Ade}, {Angil{\`e}}, {Ashton},
  {Benton}, {Devlin}, {Dober}, {Fissel}, {Fukui}, {Galitzki}, {Gandilo},
  {Hennebelle}, {Klein}, {Li}, {Korotkov}, {Martin}, {Matthews}, {Moncelsi},
  {Netterfield}, {Novak}, {Pascale}, {Poidevin}, {Santos}, {Savini}, {Scott},
  {Shariff}, {Thomas}, {Tucker}, {Tucker}, \& {Ward-Thompson}}]{Soler_2017}
{Soler}, J.~D., {Ade}, P.~A.~R., {Angil{\`e}}, F.~E., {et~al.} 2017, \aap, 603,
  A64

\bibitem[{Soler {et~al.}(2013)Soler, Hennebelle, Martin,
  Miville-Desch{\^{e}}nes, Netterfield, \& Fissel}]{Soler_2013}
Soler, J.~D., Hennebelle, P., Martin, P.~G., {et~al.} 2013, The Astrophysical
  Journal, 774, 128

\bibitem[{{Stephens} {et~al.}(2017){Stephens}, {Dunham}, {Myers}, {Pokhrel},
  {Sadavoy}, {Vorobyov}, {Tobin}, {Pineda}, {Offner}, {Lee}, {Kristensen},
  {J{\o}rgensen}, {Goodman}, {Bourke}, {Arce}, \& {Plunkett}}]{Stephens_2017}
{Stephens}, I.~W., {Dunham}, M.~M., {Myers}, P.~C., {et~al.} 2017, \apj, 846,
  16

\bibitem[{{Sternberg} {et~al.}(2014){Sternberg}, {Le Petit}, {Roueff}, \& {Le
  Bourlot}}]{Sternberg_2014}
{Sternberg}, A., {Le Petit}, F., {Roueff}, E., \& {Le Bourlot}, J. 2014, \apj,
  790, 10

\bibitem[{{Tahani} {et~al.}(2022){Tahani}, {Glover}, {Lupypciw}, {West},
  {Kothes}, {Plume}, {Inutsuka}, {Lee}, {Grenier}, {Knee}, {Brown}, {Doi},
  {Robishaw}, \& {Haverkorn}}]{Tahini_2022}
{Tahani}, M., {Glover}, J., {Lupypciw}, W., {et~al.} 2022, \aap, 660, L7

\bibitem[{{Tassis} {et~al.}(2009){Tassis}, {Dowell}, {Hildebrand}, {Kirby}, \&
  {Vaillancourt}}]{Tassis_2009}
{Tassis}, K., {Dowell}, C.~D., {Hildebrand}, R.~H., {Kirby}, L., \&
  {Vaillancourt}, J.~E. 2009, \mnras, 399, 1681

\bibitem[{{Wang}(2015)}]{WangK2015book}
{Wang}, K. 2015, {The Earliest Stages of Massive Clustered Star Formation:
  Fragmentation of Infrared Dark Clouds}

\bibitem[{{Yuen} \& {Lazarian}(2017)}]{Yuen_2017}
{Yuen}, K.~H. \& {Lazarian}, A. 2017, \apjl, 837, L24

\bibitem[{{Zhang} {et~al.}(2014){Zhang}, {Qiu}, {Girart}, {Liu}, {Tang},
  {Koch}, {Li}, {Keto}, {Ho}, {Rao}, {Lai}, {Ching}, {Frau}, {Chen}, {Li},
  {Padovani}, {Bontemps}, {Csengeri}, \& {Ju{\'a}rez}}]{Zhang_2014}
{Zhang}, Q., {Qiu}, K., {Girart}, J.~M., {et~al.} 2014, \apj, 792, 116

\end{thebibliography}



\end{document}